\documentclass[longauth]{aa}
\usepackage{graphicx}
\usepackage{txfonts}
\usepackage[colorlinks=true, linkcolor=blue, citecolor=blue, filecolor=blue, urlcolor=blue]{hyperref}
\begin{document}

   \title{A new method of reconstructing images of gamma-ray telescopes applied to the LST-1 of CTAO}

\author{
K.~Abe\inst{1} \and
S.~Abe\inst{2} \and
A.~Abhishek\inst{3} \and
F.~Acero\inst{4,5} \and
A.~Aguasca-Cabot\inst{6} \and
I.~Agudo\inst{7} \and
C.~Alispach\inst{8} \and
N.~Alvarez~Crespo\inst{9} \and
D.~Ambrosino\inst{10} \and
L.~A.~Antonelli\inst{11}
% ...
}
\institute{
Department of Physics, Tokai University, 4-1-1, Kita-Kaname, Hiratsuka, Kanagawa 259-1292, Japan
\and Institute for Cosmic Ray Research, University of Tokyo, 5-1-5, Kashiwa-no-ha, Kashiwa, Chiba 277-8582, Japan
\and INFN and Università degli Studi di Siena, Dipartimento di Scienze Fisiche, della Terra e dell'Ambiente (DSFTA), Sezione di Fisica, Via Roma 56, 53100 Siena, Italy
\and Université Paris-Saclay, Université Paris Cité, CEA, CNRS, AIM, F-91191 Gif-sur-Yvette Cedex, France
\and FSLAC IRL 2009, CNRS/IAC, La Laguna, Tenerife, Spain
\and Departament de Física Quàntica i Astrofísica, Institut de Ciències del Cosmos, Universitat de Barcelona, IEEC-UB, Martí i Franquès, 1, 08028, Barcelona, Spain
\and Instituto de Astrofísica de Andalucía-CSIC, Glorieta de la Astronomía s/n, 18008, Granada, Spain
\and Department of Astronomy, University of Geneva, Chemin d'Ecogia 16, CH-1290 Versoix, Switzerland
\and IPARCOS-UCM, Instituto de Física de Partículas y del Cosmos, and EMFTEL Department, Universidad Complutense de Madrid, Plaza de Ciencias, 1. Ciudad Universitaria, 28040 Madrid, Spain
\and INFN Sezione di Napoli, Via Cintia, ed. G, 80126 Napoli, Italy
\and INAF - Osservatorio Astronomico di Roma, Via di Frascati 33, 00040, Monteporzio Catone, Italy
% ...
}

  \date{Submitted to A\&A ; accepted September 27th, 2024}
  \abstract
   {Imaging atmospheric Cherenkov telescopes (IACTs) are used to observe very high-energy photons from the ground. Gamma rays are indirectly detected through the Cherenkov light emitted by the air showers they induce. The new generation of experiments, in particular the Cherenkov Telescope Array Observatory (CTAO), sets ambitious goals for discoveries of new gamma-ray sources and precise measurements of the already discovered ones. To achieve these goals, both hardware and data analysis must employ cutting-edge techniques. This also applies to the LST-1, the first IACT built for the CTAO, which is currently taking data on the Canary island of La Palma.  }
   {This paper introduces a new event reconstruction technique for IACT data, aiming to improve the image reconstruction quality and the discrimination between the signal and the background from misidentified hadrons and electrons.}
   {The technique models the development of the extensive air shower signal, recorded as a waveform per pixel, seen by CTAO telescopes' cameras.
Model parameters are subsequently passed to random forest regressors and classifiers to extract information on the primary particle.}
   {The new reconstruction was applied to simulated data and to data from observations of the Crab Nebula performed by the LST-1.
The event reconstruction method presented here shows promising performance improvements. The angular and energy resolution, and the sensitivity, are improved by 10 to 20\% over most of the energy range. At low energy, improvements reach up to 22\%, 47\%, and 50\%, respectively.
A future extension of the method to stereoscopic analysis for telescope arrays will be the next important step.}
   {}

    \titlerunning{A new method of reconstructing images of gamma-ray telescopes applied to the LST-1 of CTAO}
    \authorrunning{K. Abe et al.}
   \maketitlenoabstract
   \makeabstract
%
\section{Introduction}

From its inception in the 1950s to today, gamma-ray astronomy has made enormous technological and scientific progress. Surveys and multiwavelength motivated observations, regularly related to source variability, have populated this highest-energy band of the photon Universe, which has the best potential to connect to the high-energy particles bombarding our atmosphere, the cosmic rays \citep{DeAngelis:2018lra}. 

Above about 300~GeV, event rates become too low to use space-based direct detection experiments, such as \textit{Fermi}-LAT~\citep{fermilat}. The low fluxes above these energies require very large effective detection areas for meaningful scientific exploitation of the signal. 
For energies above a few tens of giga-electronvolts, gamma-ray observations can be performed indirectly from the ground, as gamma rays penetrate the upper layers of the atmosphere, inducing the creation of detectable showers of particles called extensive air showers (EASs).

The superluminal charged particles produced in these air showers emit Cherenkov radiation. Imaging atmospheric Cherenkov telescopes (IACTs) in the resulting light pool collect the Cherenkov light to detect and reconstruct the EASs' primary photons with effective areas on the order of $10^5\ \text{m}^2$. The Cherenkov light is collected by a large mirror that focuses it onto a very sensitive camera, recording a short movie of the EAS development in the atmosphere.

The Crab Nebula is a very bright source, which is useful for testing and verifying new instruments and analysis techniques for astronomy at very high energies (VHEs; 100~GeV to 100~TeV). The Crab Nebula spectrum is now measured with high precision over many energy bands \citep{Amato:2021gzt} and is used as a benchmark for the verification of the performance of IACTs and other gamma-ray instruments. The higher-energy part of this spectrum is currently measured from a few tens of giga-electronvolts up to the VHE range by IACTs \citep{HESS:2019beq,Meagher:2015igh,MAGIC:2014izm,Crab_FermiHESS2024} and up to peta-electronvolt energies by EAS experiments \citep{LHAASO:2021cbz,Abeysekara:2017mjj}.  

In this paper, we introduce a new approach for the reconstruction of IACT images produced by Cherenkov light from EASs. The goal is to provide a method of improving the quality of the data analysis of any IACTs.
This method is compatible with the data model adopted by all the telescopes of the Cherenkov Telescope Array Observatory (CTAO).
The method exploits the full recorded waveforms of all camera pixels. It performs the fitting of a model composed of a spatiotemporal prediction of the light collection in the pixels. During the fit, the model is convoluted with the precise knowledge of the camera characteristics, including the single photo-electron pulse shape and the distribution of gains in the camera. The method presented here adds to the large variety of IACT analysis techniques already available. Existing methods mostly use time-integrated images, such as the ones fitting a pre-generated template of the charge images like in~\cite{2009APh....32..231D} and ~\cite{2014APh....56...26P}, or an analytic 3D model of the EAS like in~\cite{2006APh....25..195L}. A large effort toward the development of machine-learning-based approaches is also ongoing (see for example~\cite{Jacquemont:2019pw},~\cite{2022icrc.confE.730M} and ~\cite{2021APh...12902579S}, with the latter investigating the use of waveforms in a machine-learning approach).

The method introduced here was first developed for the SST-1M telescopes~\citep{Alispach:2020rqn}. In this work, it is further improved and adapted to the Large-Sized Telescope prototype (LST-1)~\citep{lstperf}, whose camera uses photo-multiplier tubes (PMTs).

The LST-1 is located at the Roque de los Muchachos observatory on the island of La Palma at an altitude of 2147 meters and has been taking data since November 2019. 
Its reflector is composed of hexagonal mirrors that combine into an effective 23~m diameter parabolic mirror, which focuses light into a camera at a focal distance of 28~m, with a field of view of 4.3~degrees in diameter. 
The camera is equipped with 1855 1.5'' PMTs (pixels) with a hollow conical light guide, each seeing about 0.1$^\circ$ of the sky. The LST-1 can detect photons with energies ranging from $\sim$20~GeV to tens of tera-electronvolts. The LST-1 is currently in the commissioning phase and takes science commissioning data on which our event reconstruction method is tested. As we are working with a single telescope, the model and reconstruction method are currently tailored for monoscopic analysis. 
The potential for a stereoscopic analysis, using two or more telescopes, will be discussed shortly.
The LST-1 analysis pipeline, simulation production, and performance are described in depth in a first performance paper~\citep{lstperf}, which provides the standard pipeline reconstruction results, to which we refer for comparison purposes of our novel reconstruction method.

This paper is organized in the following way. In Sect.~\ref{sec:model}, we first describe the LST-1 data and how their properties are reproduced by our model. In Sect.~\ref{sec:likelihood}, the definition of the likelihood function that will be maximized to fit the model to the data is provided. Section~\ref{sec:analysis} contains the description of the full analysis pipeline used with the LST-1 and of the dataset analyzed in this paper. It also validates the Monte Carlo (MC) simulations with comparison between data and MC. The performance of the method is then estimated from simulations in Sect.~\ref{sec:perf}. In Sect.~\ref{sec:crab}, the method is applied to the observations of the Crab Nebula to perform high-level analysis and the analysis results are compared with historical data. Finally, we discuss possible future developments in Sect.~\ref{sec:future} and conclusions are drawn in Sect.~\ref{sec:conclusion}. 

\section{Data and model description}
\label{sec:model}

The IACTs focus the Cherenkov light from EASs onto a camera with pixels sensitive to single photons. These pixels and corresponding readout electronics convert incoming photons into a temporally extended electronic signal with an average integrated charge proportional to the number of photons. 
For many of the implemented cameras, including the one of the LST-1, the recording of these responses as a function of time is acquired and called a waveform. In the LST-1, the waveform is composed of 40 samples recorded at a frequency of 1.024~GHz. To extend the dynamic range, while keeping excellent precision, two gains are used in the readout electronics and the gain channel that provides the best charge resolution is selected.

\begin{figure}[ht]
    \centering
    \includegraphics[width=\linewidth]{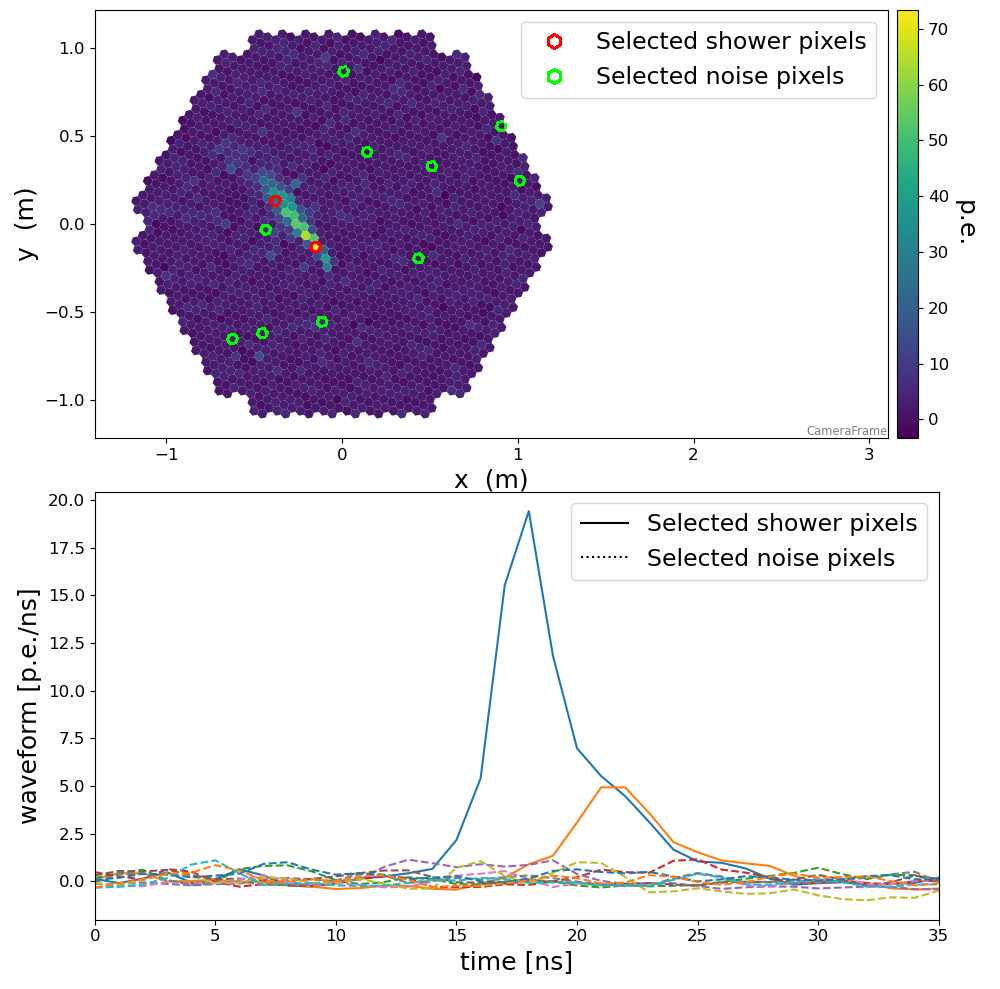}
    \caption{Example of LST-1 event. \textit{Top}: Image of the reconstructed charge for each pixel of a LST-1 event. The large majority of pixels recorded only noise. We highlight two pixels hit by the shower light and and several others without any Cherenkov signal, indicated by red and green circles, respectively.
    \textit{Bottom}: Calibrated waveforms for the selected pixels of the image at the top. }
    \label{fig:waveform}
\end{figure}

The likelihood reconstruction method that we present in this paper has been applied to calibrated waveforms.
The calibration includes pixel-wise corrections to the gain and timing, which are derived from specific calibration data. 
The baseline was subtracted and the gain factor was applied to obtain the waveform in photo-electrons per sample unit.\footnote{It is also possible to apply the method before this step by including the gain and baseline in the likelihood function as done in the original implementation~\citep{cyrilthesis}} An LST-1 event is thus a set of 1855 waveforms combining random pedestal fluctuations and the signal from the EAS. 
Examples of such waveforms are shown in Fig.\ref{fig:waveform}. The main contribution to the baseline fluctuation is the night sky background (NSB). 
The waveforms were synchronized using independently measured time-shift corrections on the relative timing between pixels.

\begin{figure*}
    \centering
    \includegraphics[width=\textwidth]{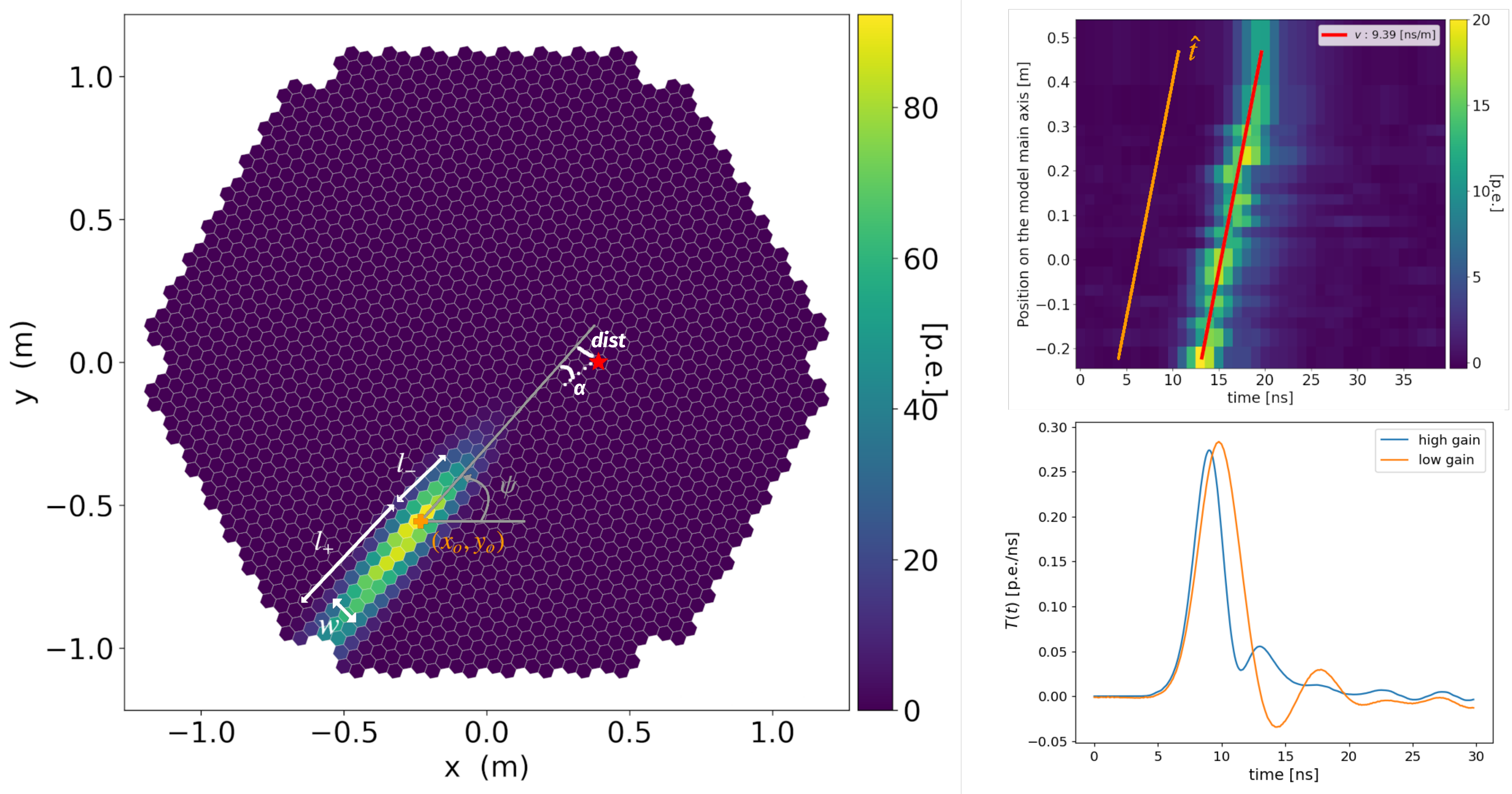}
    \caption{Model description. \textit{Left}: 2D asymmetric Gaussian spatial model obtained after fitting the full model to an MC gamma-ray event. The red star is the position of the gamma-ray source in the camera. Spatial model parameters, and source-dependent analysis parameters ($\alpha$ and $dist$), are also shown. \textit{Top right}: Waveform amplitude distribution as a function of time and of the position along the fit main axis in the same event. The orange line represents the linear shift between the time of arrival of the signal at different positions along the shower main axis given by the temporal model. The red line is the same temporal model shifted to the maximum of the waveforms for illustration. \textit{Bottom right}: Template of the normalized pulsed response of a pixel to a single photo-electron in the two gain channels used by LST-1.}
    \label{fig:model}
\end{figure*}

The method presented here models the development of a gamma-ray-initiated electromagnetic EAS in the photo-detection plane of the camera.
The event characteristics, predicted by the model, are compared to the event's waveforms. The best-fit parameters of the model correspond to those maximizing the likelihood of the model for the event.
This model must adhere to a set of key requirements:
\begin{itemize}
    \item it must predict the number of photons reaching each pixel and the associated timing;
    \item it must include the pixel response;
    \item it must be simple enough to enable a quick convergence of the fit;
    \item it must be accurate enough to improve the reconstruction of the primary particle properties.
\end{itemize}
Electromagnetic EASs develop around the primary particle trajectory, and Cherenkov emission occurs in the region of the EAS where the energetic electrons and positrons are. 
The emitted light is registered when the shower produces a number of photo-electrons in the camera above the trigger threshold. 
The shower light, focused by the telescope mirror, forms a roughly elliptical image with a distribution of photo-electrons decreasing toward its edges. Therefore, we decided to model the spatial distribution of charge using a 2D Gaussian. Moreover, the charge distribution exhibits an asymmetry along the longer axis of the image~\citep{Fegan:1997db}, which we included in the model. This asymmetry is due to the fact that the most energetic particles in the EAS are located close to the point of interaction.
The spatial model is ultimately characterized by a set of seven parameters: the total number of photo-electrons, $N$, the position of the center of the model in the camera frame ($x_o$, $y_o$), the two Gaussian standard deviations along its main axis on each side of the maximum and the one along the secondary axis ($l_+$, $l_-$ and $w$), and the angle, $\psi$, between the shower main axis and the camera x axis.

\begin{equation}
    \mu(x,y) = \frac{N}{\pi (l_+ + l_-)w} exp(\frac{-L^2}{2l_\pm^2}) exp(\frac{-W^2}{2w^2})
,\end{equation}
with
\begin{equation}
    \begin{array}{c} L = (x-x_0)cos(\psi) + (y-y_0)sin(\psi) \\
    W = (y-y_0)cos(\psi) - (x-x_0)sin(\psi) 
    \end{array}
,\end{equation}
and where $l_\pm$ is $l_+$ or $l_-$ depending on the sign of $L$.
This spatial component of the model gives the expected number of photo-electrons, $\mu$, in each pixel, as is illustrated for a simulated gamma-ray event in Fig.\ref{fig:model}-\textit{left}, where the spatial model parameters are also shown.

The evolution of the time of arrival of the light as a function of the position of emission is directed by the EAS extension in the atmosphere and the velocity of the emitted Cherenkov light. The resulting time profile is strongly dependent on the impact parameter; that is, the distance between the telescope and the EAS axis of the shower, as is illustrated in \cite{2008ICRC....5.1253M, 2009APh....30..293A}. Most EASs have a large impact parameter, in which case the position of the center of gravity of the EAS light in the camera moves at a constant speed along the main shower axis, the projection of the development of the shower in the atmosphere. Therefore, we applied a linear temporal model to describe the development of the image in the camera plane as a function of the position of the pixel in the camera projected onto the spatial main axis. Due to the higher velocity of particles compared to the velocity of light in the atmosphere, the time difference between the arrival of photons emitted early and late in the shower development reduces with the impact parameter, reaching zero at intermediate impacts. Using our gamma-ray application MC simulation,\footnote{As is defined in Sect. \ref{sec:analysis_desc}, and weighted as in Sect. \ref{sec:datamc}} we observe this happening at impact parameters between 100 and 125 meters, decreasing with energy. The fraction of events with impact parameters larger than 125 meters is 50\%, 78\%, and 90\% for the energy ranges [10~GeV$-$100~GeV], [100~GeV$-$1~TeV], and [1~TeV$-$10~TeV], respectively. In cases of very low impacts, the photons will arrive first near the center of the image and then at the edges. Still, the linear time gradient carries relevant information on the shower and can thus be used in the analysis. The use of a more complex and realistic temporal profile is not covered in this work. 
Our linear temporal model is parametrized by the time gradient, $v$, representing the time shift per unit distance along the main axis of the shower, and a reference time, $t_o$, for the position ($x_o$, $y_o$). It provides $\hat{t}$, a reference time per pixel for the Cherenkov photons' time of arrival. This is illustrated in Fig.\ref{fig:model}-\textit{top-right}, representing the distribution of the sum of waveform amplitudes as a function of time and the projection of the pixel position on the main axis of the spatial model component.
No dispersion of the arrival time in a single pixel is included as this model proved to already be a good approximation with the sampling rate used here. 

The last component of the model is a pixel response function. 
It represents the waveform induced by the detection of photons in a pixel. 
This includes the light sensor, along with the response from the front-end electronics. 
Consequently, the response of the pixel to $X$ photo-electrons can be calculated as a linear combination of the normalized single photo-electron responses. 
We indicate with $T(t)$ the normalized pulsed response to a single photo-electron as a function of time. Since we are neglecting the time dispersion of the photon arrival within a single pixel, the response of a pixel to $X$ photo-electrons reduces to $X\times T(t)$, simply scaling the model waveform. 
Since two gain channels are available in LST-1, two associated pulse templates are provided and used accordingly. They are shown in Fig.\ref{fig:model}-\textit{bottom-right}. 
The temporal model gives the time corresponding to the arbitrary zero of the single photo-electron response template. Consequently, $\hat{t}$ is shifted compared to the times of the maximum of the waveforms, as is visible in Fig.\ref{fig:model}-\textit{top-right}.

\section{Definition of the model likelihood}
\label{sec:likelihood}

The complete likelihood of the model was estimated for the event waveform. The waveform is a set of signal values, $S_{ij}$, for each pixel, $i$, and each sample of time, $j$. The full likelihood of the model is the product of the likelihood of each sample, $\mathcal{L}_{ij}$. To reconstruct the model parameters, we need to maximize the log-likelihood:

\begin{equation}
\ln \mathcal{L} = \sum_i^{pixels}\sum_j^{times} ln \mathcal{L}_{ij}
.\end{equation}

The single sample likelihood is represented by the probability of observing the signal, $S_{ij}$, knowing $\mu_i$, the average number of photo-electrons in the pixel, $i$, from the spatial component of our model, $T_i$, the normalized single photo-electron response template for the gain used in the pixel, $i$, and $\hat{t_i}$, its reference time from the temporal component of our model.
Three effects need to be taken into account. 
First, the exact distribution of Cherenkov light emission by the EAS particles and the conversion of photons to photo-electrons by PMTs are stochastic. 
Consequently, the probability mass function of receiving $k$ photo-electrons in the pixel, $i$, knowing $\mu_i$ is a Poisson law:\footnote{Originally, the method was developed to be compatible with pixels using Silicon Photo-multipliers, so crosstalk was also taken into account and a generalized Poisson law~\citep{2012NIMPA.695..247V} was used.}

\begin{equation}
    P = \mathcal{P}(k|\mu_i) = \frac{\mu_i^{k}}{k!}e^{-\mu_i}
.\end{equation}

\begin{figure}[ht]
    \centering
    \includegraphics[width=\linewidth]{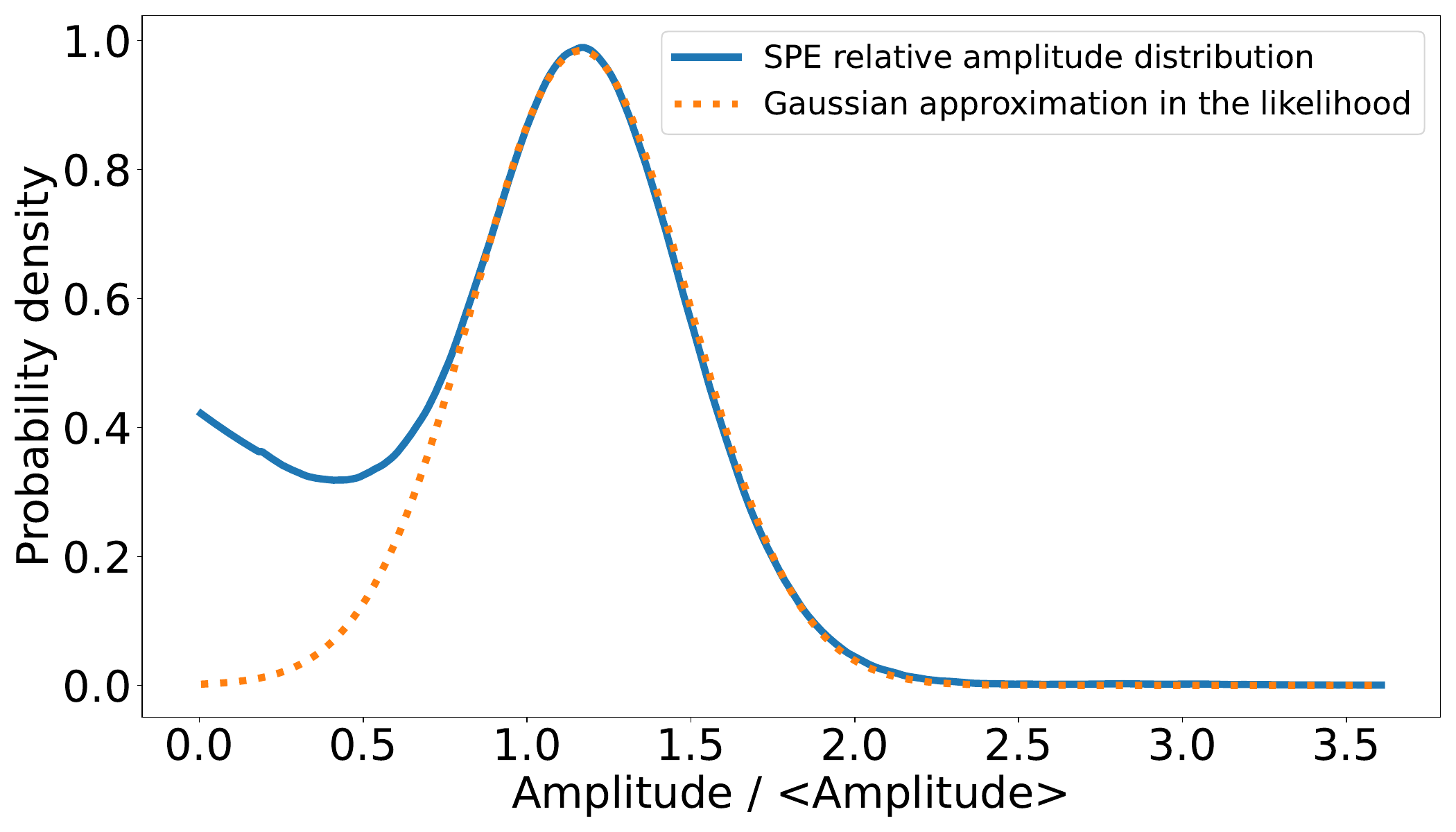}
    \caption{Single photo-electron (SPE) amplitude distribution and Gaussian model used to approximate its variance. The SPE amplitude is given relative to the average amplitude of the signal produced by a single photon converted in a PMT.}
    \label{fig:spe}
\end{figure}

Second, the normalization of the response of the pixel to any photo-electron is randomly distributed. 
It is illustrated, for the case of LST-1, in Fig.\ref{fig:spe}. In the likelihood computation, we approximate this distribution by the Gaussian also shown in Fig.\ref{fig:spe} with the gain smearing, $\sigma_s$, as the standard deviation. 
Finally, the baseline of the waveform fluctuates from NSB and electronic noise. 
The baseline fluctuations come from a large number of effects and are mostly represented by a Gaussian probability density function with a standard deviation, $\sigma_e$. In PMTs, afterpulses lead to a small deviation from the Gaussian behavior, which is not accounted for in the following likelihood. All Gaussian terms (one for the baseline and one for each photo-electron) can be combined in a single Gaussian.
It represents the probability of observing a signal, $S_{ij}$, from $k$ photo-electrons. 
We denote the time associated with $S_{ij}$ as $t_{ij}$. In this case, the expected charge for this sample is $k \times T_i(t_{ij}-\hat{t_i}$). 
We have:

\begin{eqnarray}
G &=& \mathcal{P}(S_{ij}|k, t_{ij} - \hat{t_i}, T_i)\\
G &=& \frac{1}{\sqrt{2\pi}\sigma_k}\exp{\left(-\frac{(S_{ij} - k T_i(t_{ij} - \hat{t_i}))^2}{2\sigma_k^2}\right)}
.\end{eqnarray} 

Here, we have introduced $\sigma_k = \sqrt{\sigma_e^2 + k(\sigma_s T_i(t_{ij} - \hat{t_i}))^2}$ as the standard deviation of the combined Gaussian. The total probability of observing $S_{ij}$ from our model is then a sum of the contributions of all possible numbers of photo-electrons, $k \in [0,\infty]$:

\begin{eqnarray}
    \mathcal{L}_{ij} &=& \mathcal{P}(S_{ij}|\mu_i, t_{ij} - \hat{t_i}, T_i)\\
    &=& \sum_{k=0}^\infty \mathcal{P}(k|\mu_i) \mathcal{P}(S_{ij}|k, t_{ij} - \hat{t_i}, T_i)\\
    &=& \sum_{k=0}^\infty P \times G
\end{eqnarray}

\begin{equation}
    \mathcal{L}_{ij} = \sum_{k=0}^\infty \frac{\mu_i^{k}}{k!}e^{-\mu_i} \times \frac{1}{\sqrt{2\pi}\sigma_k}\exp{\left(\!-\frac{(S_{ij}\! -\! k T_i(t_{ij}\!-\!\hat{t_i}))^2}{2\sigma_k^2}\right)}
.\end{equation}

The likelihood function contains an infinite sum of computationally expensive terms. Therefore, two approximations were implemented. First, the likelihood converges to a fully Gaussian function when the signal increases~\citep{cyrilthesis}. Hence, we introduced a transition charge, $\mu_{trans}$, such that pixels with $\mu_i > \mu_{trans}$ use the following Gaussian approximation:

\begin{equation}
    \mathcal{L}_{ij} =\frac{1}{\sqrt{2\pi}\sigma_{\mu i}}\exp{\left(\!-\frac{(S_{ij}\! -\! \mu_i T_i(t_{ij}\! -\! \hat{t_i}))^2}{2\sigma_{\mu i}^2}\right)}
\end{equation}
With $\sigma_{\mu i} = \sqrt{\sigma_e^2 + \mu_i (T_i(t_{ij} - \hat{t_i}))^2}$.

\begin{figure}
    \centering
    \includegraphics[width=\linewidth]{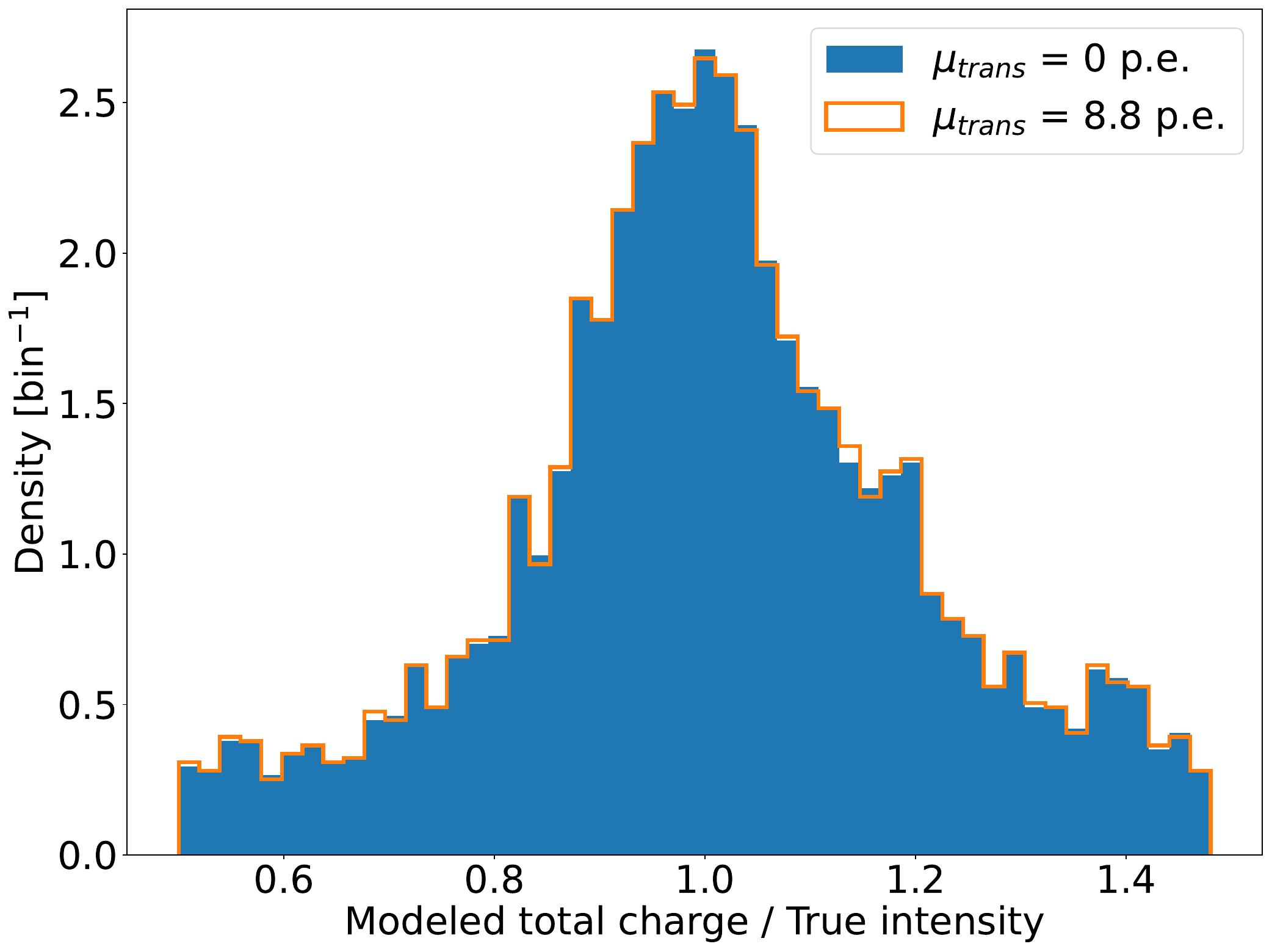}
    \caption{Distribution of the ratio of total charge from the likelihood fit of our model divided by the true number of photo-electrons simulated in the event. Distributions are very similar when using only the Gaussian approximation of the likelihood for all pixels, and when using the complete likelihood function (with $k_{max}$ = 20) for pixels with an expected charge of less than 8.8 p.e.}
    \label{fig:truevsfit_intensity}
\end{figure}

The second approximation is to limit the infinite sum in $\mathcal{L}_{ij}$ to a maximum $k_{max}$. It must be selected so that the terms of the sum with $k > k_{max}$ are negligible. $\mu_{trans}$ is adapted to $k_{max}$ to guarantee this behavior when the Gaussian approximation is not used. The value of $k_{max}$ is configurable but can be constrained by software limitations (e.g., the maximum factorial usable with a 64-bit integer is 20). The current configuration for analysis of LST-1 mono data uses $\mu_{trans}= 0$, meaning that all pixels are processed using the Gaussian approximation. It was verified on MC simulations that such a configuration has nearly no effect on analysis performance compared to using higher possible values of $\mu_{trans}$, while the required computational power is significantly reduced. This is illustrated in Fig.\ref{fig:truevsfit_intensity}, where the ratio of the total fit charge from our model divided by the true number of photo-electrons from the simulation is shown for two configurations. The case using $\mu_{trans}=0$~p.e. is compared to the case using $\mu_{trans}\approx8.8$~p.e., the latter being associated with $k_{max} = 20$.\footnote{This requires that the $P(k>k_{max})$ terms be less than $(1/k_{max})\%$, which should allow us to ignore less than 1\% of the Poisson probability mass function}

A preselection of pixels and times was also performed to avoid wasting resources on regions of the data far away from the signal. It can also limit the number of stars in the fit region, and thus limit the number of pixels with complex behaviors. Indeed, stars add light in specific pixels, increasing their waveform fluctuations. In the case of bright stars, it can also lead to an automatic adjustment of the pixels gains. Current MC simulations do not account for such localized and time-dependent effects. Only pixels contained in an ellipse defined from Hillas' parameters~\citep{hillas_parameters} with three times its semi-major and minor axes were used. This choice was not optimized for analysis or computing performance but should keep all signal pixels for gamma-ray events.

\section{Analysis}
\label{sec:analysis}
\subsection{Pipeline and data description}
\label{sec:analysis_desc}

The method described here was implemented in the \textit{cta-lstchain} pipeline~\citep{lstchain_zen} as an alternative to image reconstruction based on the extraction of Hillas' parameters. Usage of the latter for LST-1 is covered in~\citep{lstperf}. \textit{cta-lstchain} is the analysis pipeline developed to analyze LST-1 data until the CTAO data analysis pipeline is released. It performs the analysis of LST-1 data and transforms raw waveforms into a collection of reconstructed gamma-like events. The standard event processing follows the steps: 1.~waveform calibration, 2.~charge and peak time extraction, 3.~image cleaning, 4.~Hillas parametrization, 5.~primary particles property inference, and 6.~event selection and instrument response function (IRF) creation. Hillas parametrization consists of the extraction of the image momenta from the integrated charge images\footnote{Obtained using a $LocalPeakWindowSum$ charge extraction algorithm~\citep{local_peak_window_sum}} of IACTs. It has been shown to be a simple and robust way to extract useful information from the Cherenkov telescope data.

Our method, which we label as “LH~fit,” works using the calibrated waveforms to perform an image parametrization in the place of steps 2, 3, and 4 described above. 
It then replaces the Hillas parametrization used in the primary particle properties inference (step 5) with our model parameters. 
The fit was initialized using seed parameters derived from Hillas' image parametrization. The fit was made by minimizing $-2ln\mathcal{L}$ with \textit{iminuit}~\citep{iminuit}.

\begin{figure}[ht]
  \centering
  \includegraphics[trim=0 0 2 0,clip, width=\linewidth]{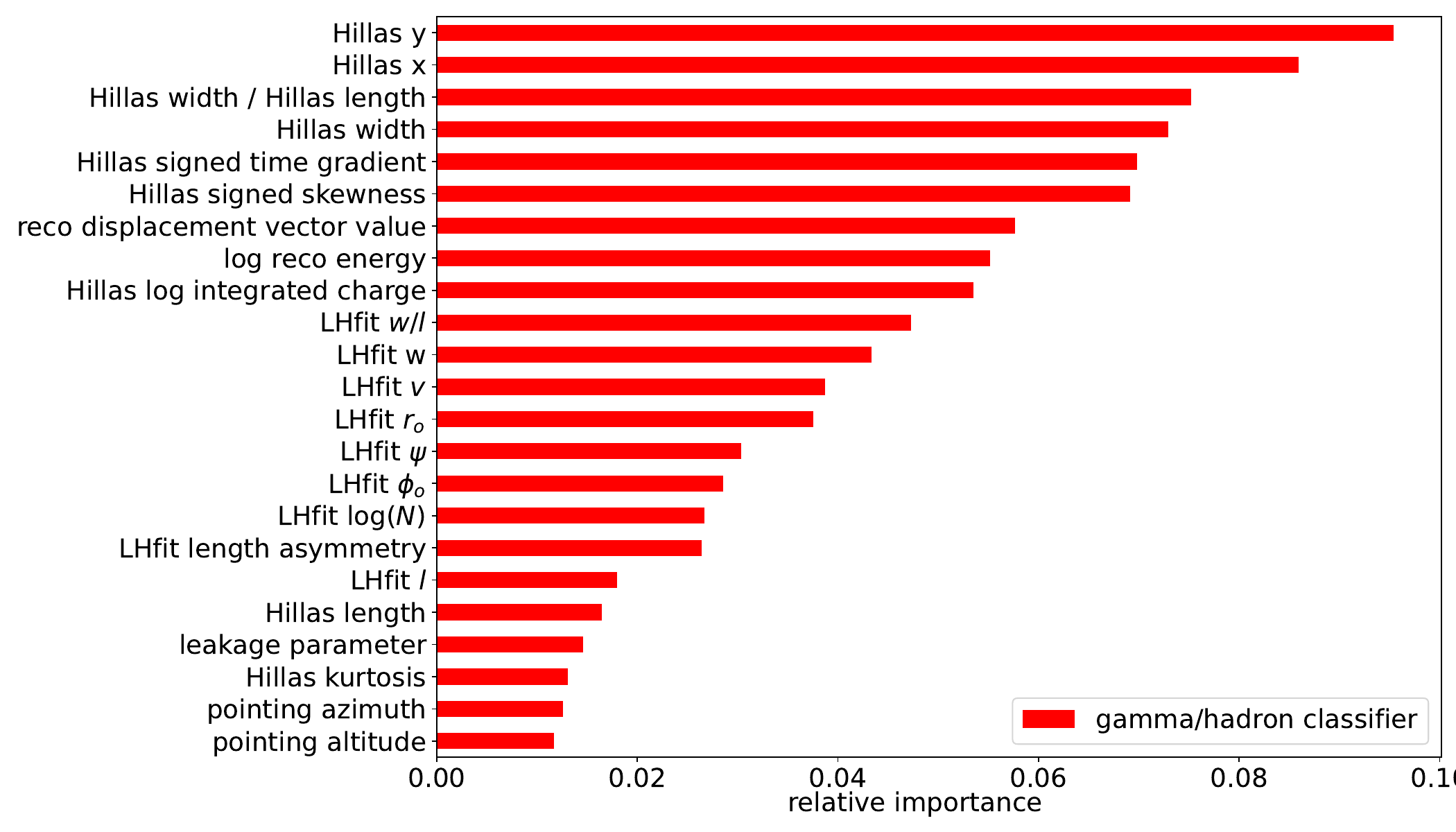}
  \caption{Relative importance of the features of our gamma-hadron classifier. Parameters labeled LHfit are derived from our model. Parameters labeled Hillas are Hillas' parameters. The classification is dominated by Hillas' parameters, with in particular the ratio of Hillas' width over length being the most important after the centroid position. The importance of this parameter was expected, since hadronic EASs are generally wider than electromagnetic EASs.}
  \label{fig:rf4-class}
\end{figure}

\begin{figure}[ht]
  \centering
  \includegraphics[width=\linewidth]{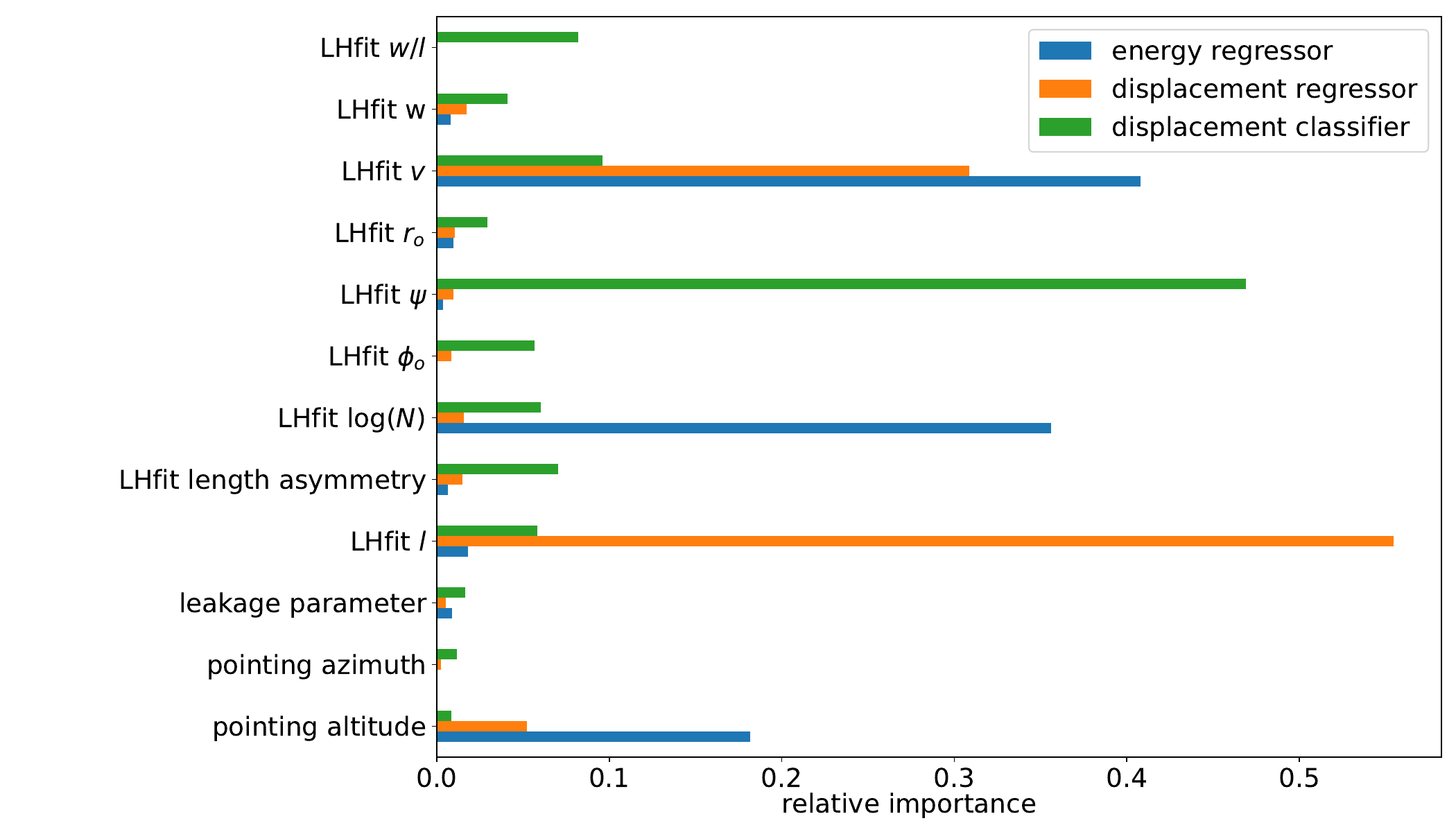}
  \caption{Relative importance of the features of our energy and direction RFs. The energy regression is mostly related to the total light of the fit model and to the temporal development that indirectly relates to the impact parameter, and thus the distance between the telescope and the EAS. The displacement regressor, which gives the angular separation between the source and the image centroid, has a strong dependence on the model length and temporal development. Finally, the displacement classifier, determining which side of the image centroid the source is located on, is largely dominated by the LH~fit, $\Psi$, which combines information on the orientation of the model and the direction of temporal development.}
  \label{fig:rf4-noclass}
\end{figure}

After the extraction of the model parameters, the energy, direction of arrival, and gamma-hadron classification score (called gammaness) of each event were estimated using random forests (RFs) trained on simulated data. In total, four RFs were used: a regressor for the energy reconstruction, a regressor for the value of the displacement vector between the EAS signal core and the source position, a classifier for the vector orientation, and a classifier for the gamma-hadron classification. The package used for this purpose is \textit{SciPy}~\citep{2020SciPy-NMeth}. The parameters used for the RF were (depending on reconstructed quantity, see Fig.\ref{fig:rf4-class}-\ref{fig:rf4-noclass}):
\begin{itemize}
    \item log $N$, the total charge of the modeled image on a log10 scale;
    \item $r_o$ and $\phi_o$, the circular coordinate representation of the center of the spatial model ($x_o$, $y_o$);
    \item the average model length ($l = (l_+ + l_-)/2$) and the associated length asymmetry parameter ($\pm l_+/l_-$), where the sign depends on whether the longer side is the early or late part of the signal development;
    \item the model width, $w$, and the ratio, $w/l$;
    \item $\psi$,  the angle between the shower main axis and the camera x axis;
    \item $v$, the time gradient in the temporal model;
    \item a leakage parameter, defined as the fraction of charge in pixels surviving cleaning located in the last two layers of pixels at the edge of the camera. This parameter was defined using the standard charge extraction and cleaning;
    \item the telescope pointing information: azimuth and altitude angles;
    \item the reconstructed energy (log scale) and value of the reconstructed displacement vector. These were only used for the gamma-hadron classification;
    \item for the gamma-hadron classification, the parameters extracted through the model alone are less effective than the standard Hillas' parameters. We thus included fit and Hillas' parameters (described in~\cite{lstperf}) in the RF features.
\end{itemize}


The high-level analysis of the data reduced with \textit{cta-lstchain} was finally performed with the package \textit{gammapy} version 1.0.1 \citep{2023A&A...678A.157D, gammapy101}, a package dedicated to the high-level analysis of astronomical data. This paper uses the same three datasets as in \citep{lstperf}: a set of MC simulations was used to train the RFs (training MC), another set of MC simulations was used to check the agreement between real observation data and MC data as well as to produce the IRFs for the data analysis (application MC), and observations of the Crab Nebula were also used.

\begin{figure}[ht]
    \centering
    \includegraphics[width=\linewidth]{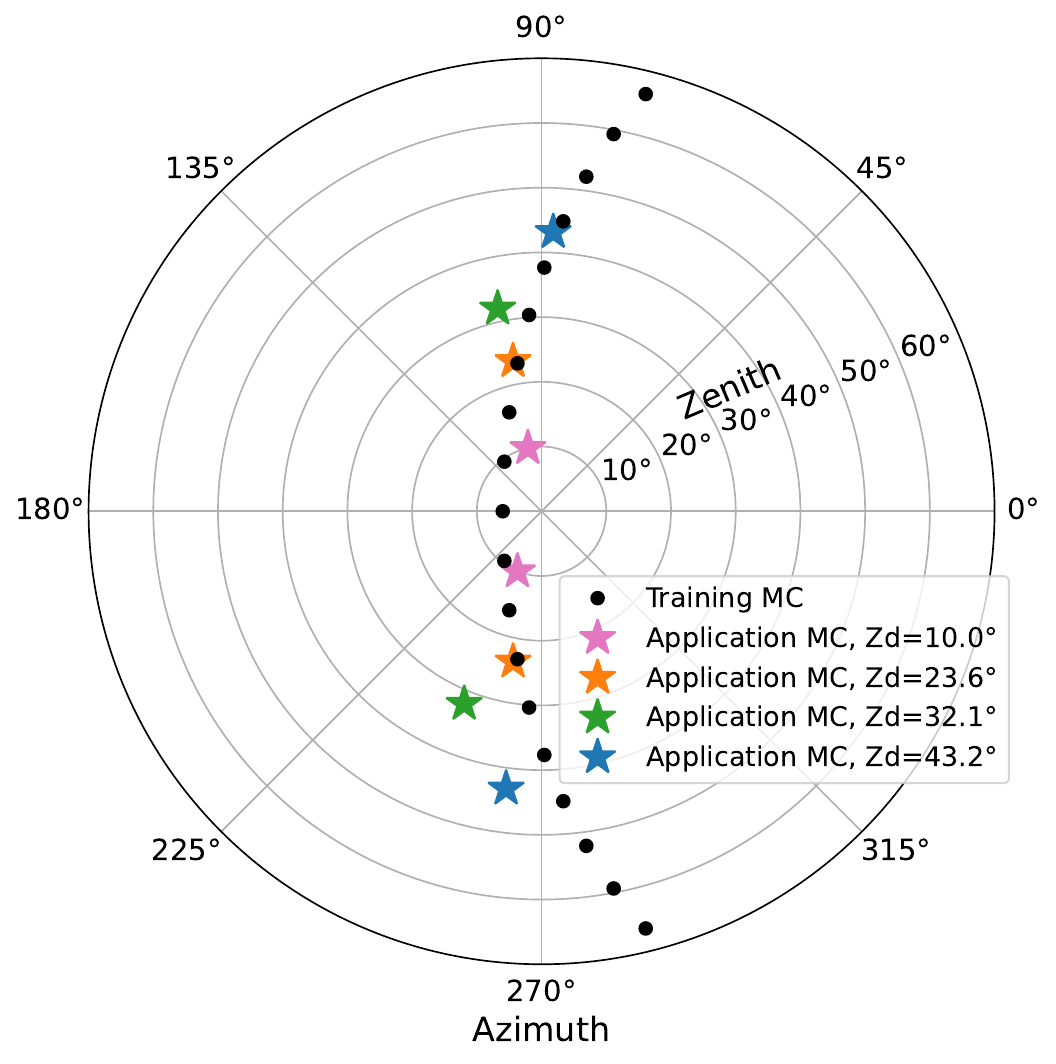}
    \caption{Position of the pointings in the simulation productions used in this paper. Zd, for zenith distance, is the angle between the zenith and the pointing position. The black points are for our training MC set, produced along the trajectory of the Crab Nebula. The stars are the pointings of the application MC sets.}
    \label{fig:MCpointings}
\end{figure}

The training MC set was simulated at pointings following the declination of the Crab Nebula (see the black points in Fig.~\ref{fig:MCpointings}). It contains both diffuse gamma rays and proton simulations. 
Only gamma-ray simulations were used for the training of the energy and direction reconstruction, while both gamma-rays and protons were used to train the gamma-hadron classifier.
The application MC simulations were used to evaluate analysis performance and to create IRFs. 
The IRFs currently in use are the energy migration matrix, which links the energies reconstructed by the RF to the true energy of the events, and the effective area of the instrument, which is used to convert the observed number of excess events to fluxes. 
The application MC simulations were divided into eight pointings near the Crab Nebula path at 10, 23, 32, and 43 degrees from the zenith with two azimuth angles each (see Fig.~\ref{fig:MCpointings} stars). The NSB level in both MC sets was adjusted, in the events waveforms, to the level observed in the Crab Nebula field of view.

The Crab Nebula dataset corresponds to a total of 36 hours of observations taken between November 2020 and March 2022.
\newline

\noindent\textbf{\textit{Source-dependent analysis}}

It is possible to add a set of parameters accounting for the known source position in the camera plane. This technique, already used with Hillas' parametrization, can also be used with our method. In our case, the parameters of interest are:
\begin{itemize}
    \item $\alpha$, the angle between the longer axis of the model and the line connecting the centroid of the model and the position of the source;
    \item $dist$, the distance between the ($x_0$, $y_0$) of the model and the position of the source.
\end{itemize}

The results of our pipeline using this slightly different analysis are also shown in the following sections. No direction reconstruction was performed in this case, as it is assumed to be known.

\subsection{Comparison between observed and simulated data}
\label{sec:datamc}

\begin{figure}[ht]
    \centering
    \includegraphics[width=\linewidth]{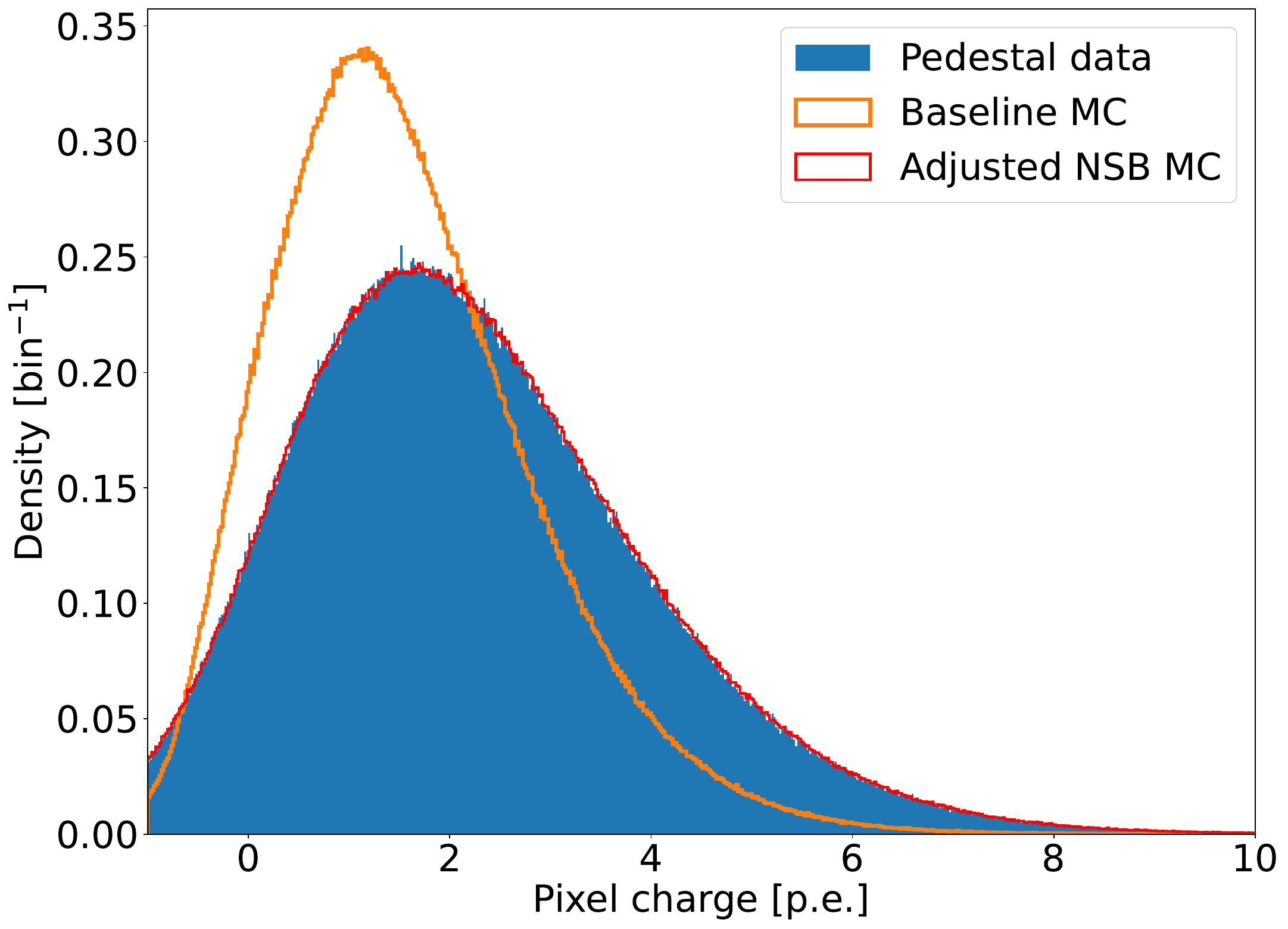}
    \caption{Distribution of pixel charges for data without EAS contribution. Pedestal events, taken during standard data taking without triggers based on EAS detection, were used for real observations. For MC, pixels with a true charge of 0~p.e. from Cherenkov photons were considered. A significant improvement of the agreement between data and MC is observed when adjusting the NSB level.}
    \label{fig:NSBtuning}
\end{figure}

Prior to the evaluation of the method's performance, we needed to ensure that our simulation correctly reproduces the observation data. 
To do so, we compared the basic quantities' distributions, such as the individual pixels' charge distributions and the distribution of image intensity. 
Intensity refers to the total charge extracted in pixels surviving cleaning in the standard event processing (steps 2 and 3). Figure~\ref{fig:NSBtuning} shows the individual pixel charge distribution with no EAS contribution. 
The MC with an adjusted NSB shows a very similar distribution when compared to the data. The NSB adjustment was performed by injecting single photo-electron pulses directly into the waveforms. This differs from \citep{lstperf}, for which an adjustment of the integrated charge per pixel was done. The NSB adjustment does not include localized effects from stars, which are responsible for brighter pixels than expected.
Then, the first step in evaluating the method was to assess the agreement between the observed data and the simulation for model parameters from our parametrization and the outputs of the RFs. We applied a loose preselection of events to reduce the statistical fluctuations of the background contribution and systematic effects from the evolution of data-taking conditions: 
\begin{itemize}
    \item only events with an intensity above 80 p.e. were considered;
    \item an upper limit on the angular distance between the true and reconstructed source direction ($\theta$).
\end{itemize}
In observation data, the same selection was applied to a region in the sky, a so-called OFF region, which is symmetric compared to the source position with respect to the telescope pointing direction. 
The background distribution extracted in this way was used to quantify the contribution from the excess signal in the data. 
This remaining excess in observation was then compared to the gamma rays from our application MC after normalization of the number of events following the expected source spectral energy distribution (SED). The Crab Nebula SED is very well known and stable in the energy band where IACTs are sensitive~\citep{2015JHEAp...5...30A}.

\begin{figure*}
    \centering
    \includegraphics[width=\textwidth]{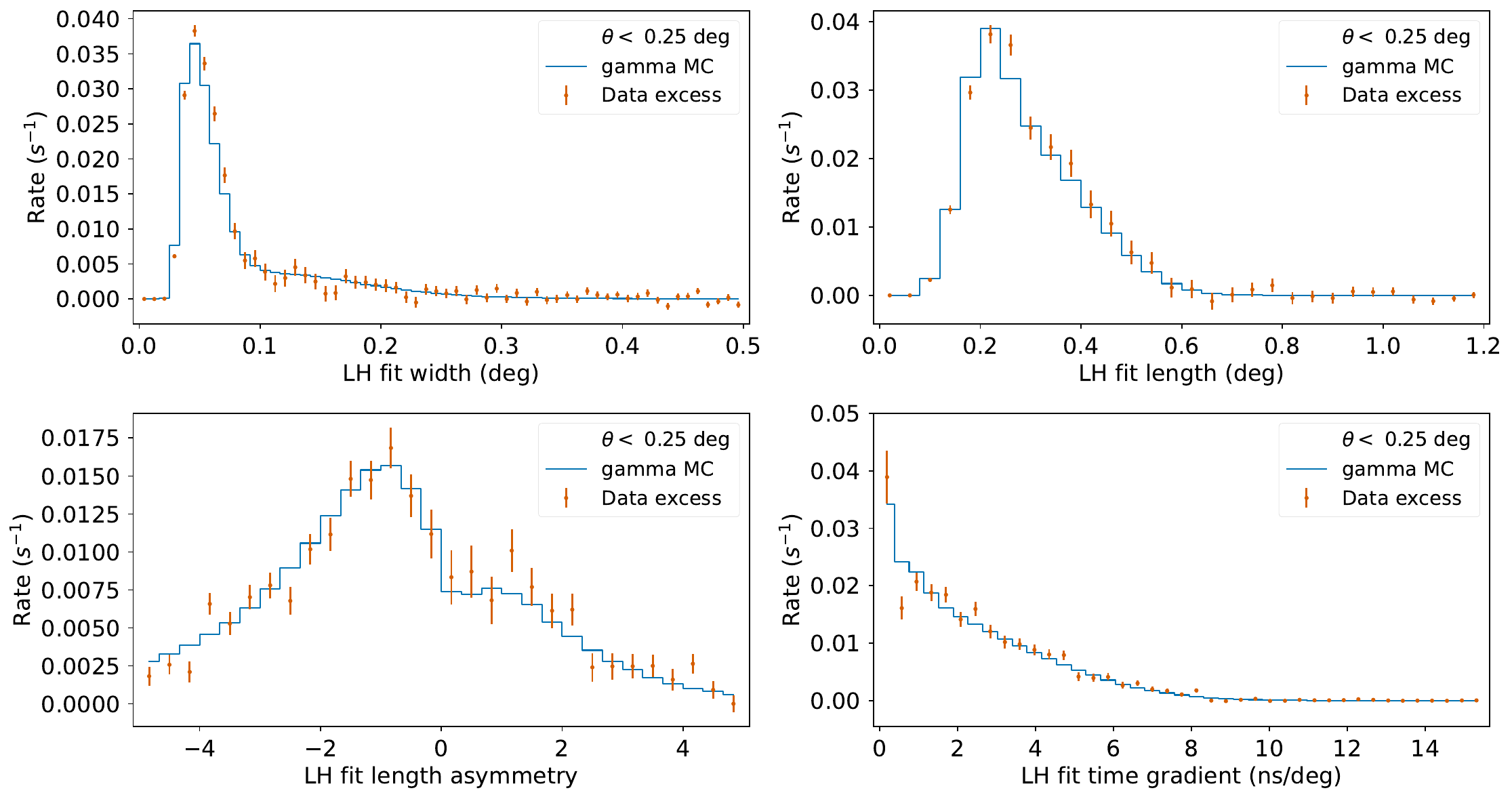}
    \caption{Comparison of the model parameters' distribution between excess events from Crab Nebula observation and simulated gamma events with an energy distribution following the Crab Nebula spectrum. Four model parameters' distribution for image intensities between 800 and 3200 p.e. are shown.}
    \label{fig:model_param_datamc}
\end{figure*}

\begin{figure*}
    \centering
    \includegraphics[width=\textwidth]{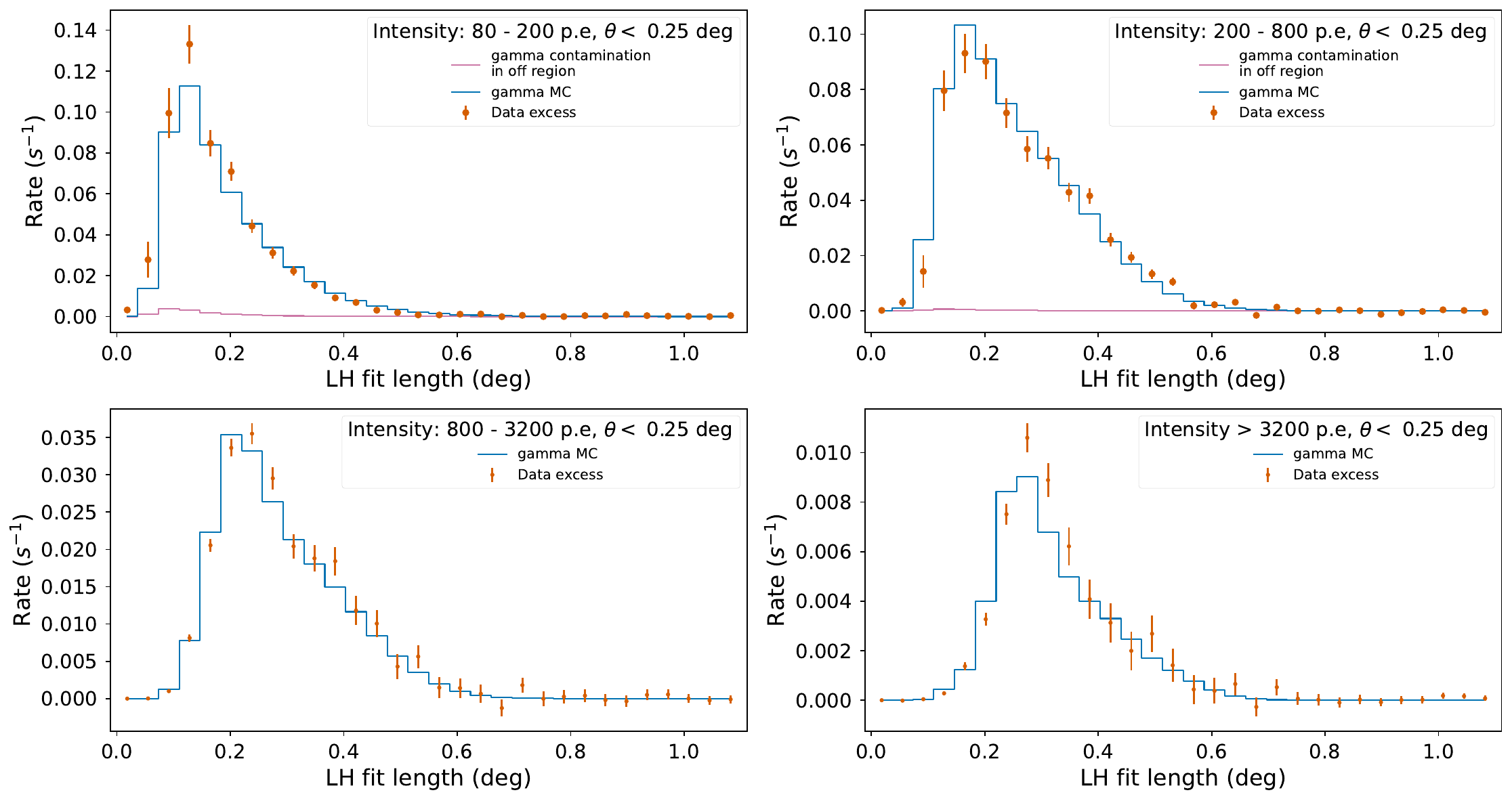}
    \caption{Same as Fig.\ref{fig:model_param_datamc} but showing only the LH~fit length parameter for four image intensity ranges. Using these four intensity ranges allows us to see the evolution of the agreement between data and MC for different primary energy and signal-to-noise ratios in the pixels.}
    \label{fig:model_param_datamc_length}
\end{figure*}

\begin{figure*}
    \centering
    \includegraphics[width=\textwidth]{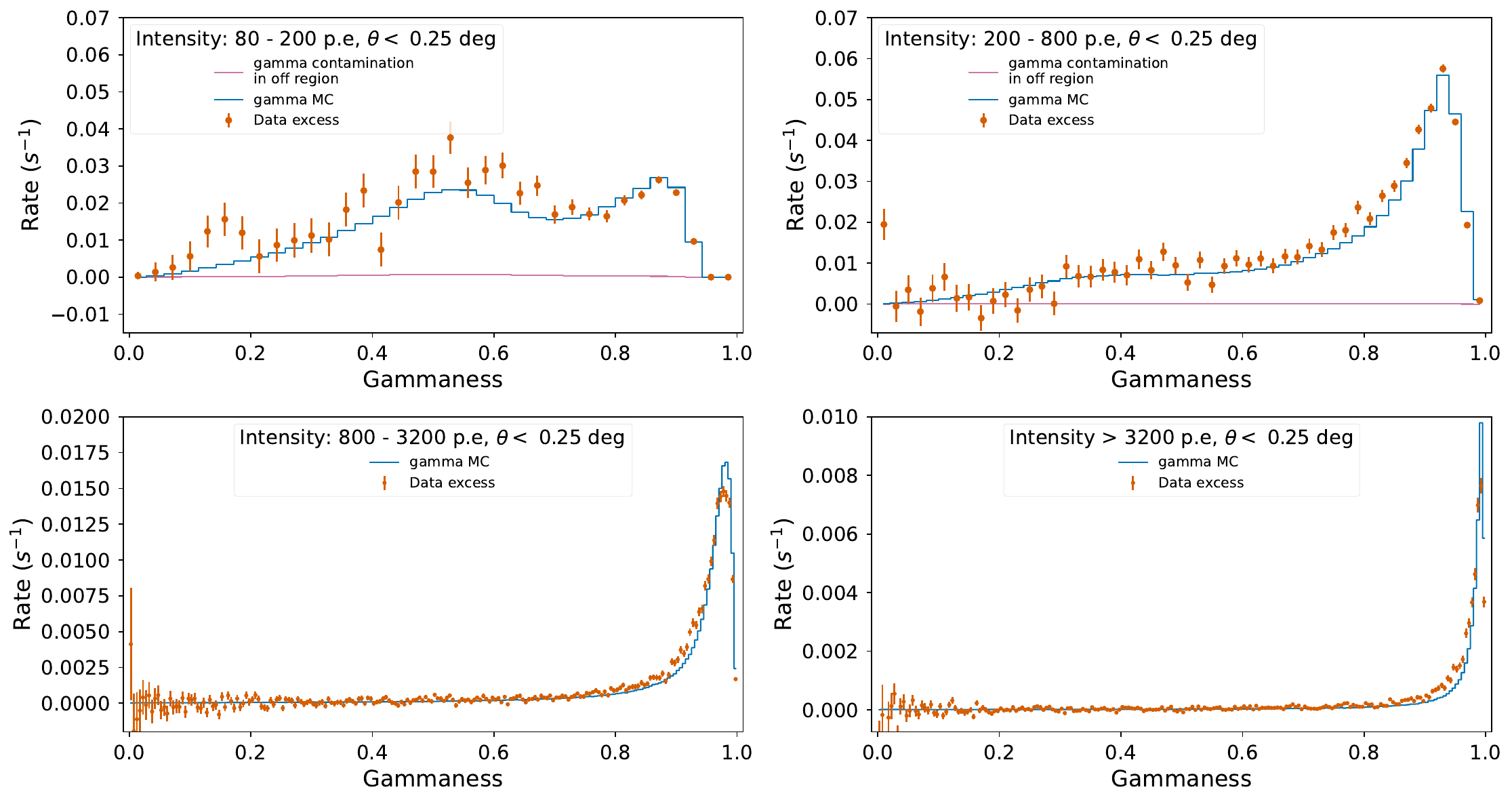}
    \caption{Comparison of the gammaness distribution between excess events from Crab Nebula observation and simulated gamma events with an energy distribution following the Crab Nebula spectrum. A comparison is made for four image intensity ranges. The distribution shifts closer to one with higher intensity, showing the expected improvement of the gamma-hadron discrimination power with image intensity.}
    \label{fig:hl_param_datamc_gammaness}
\end{figure*}

\begin{figure}[ht]
    \centering
    \includegraphics[width=\linewidth]{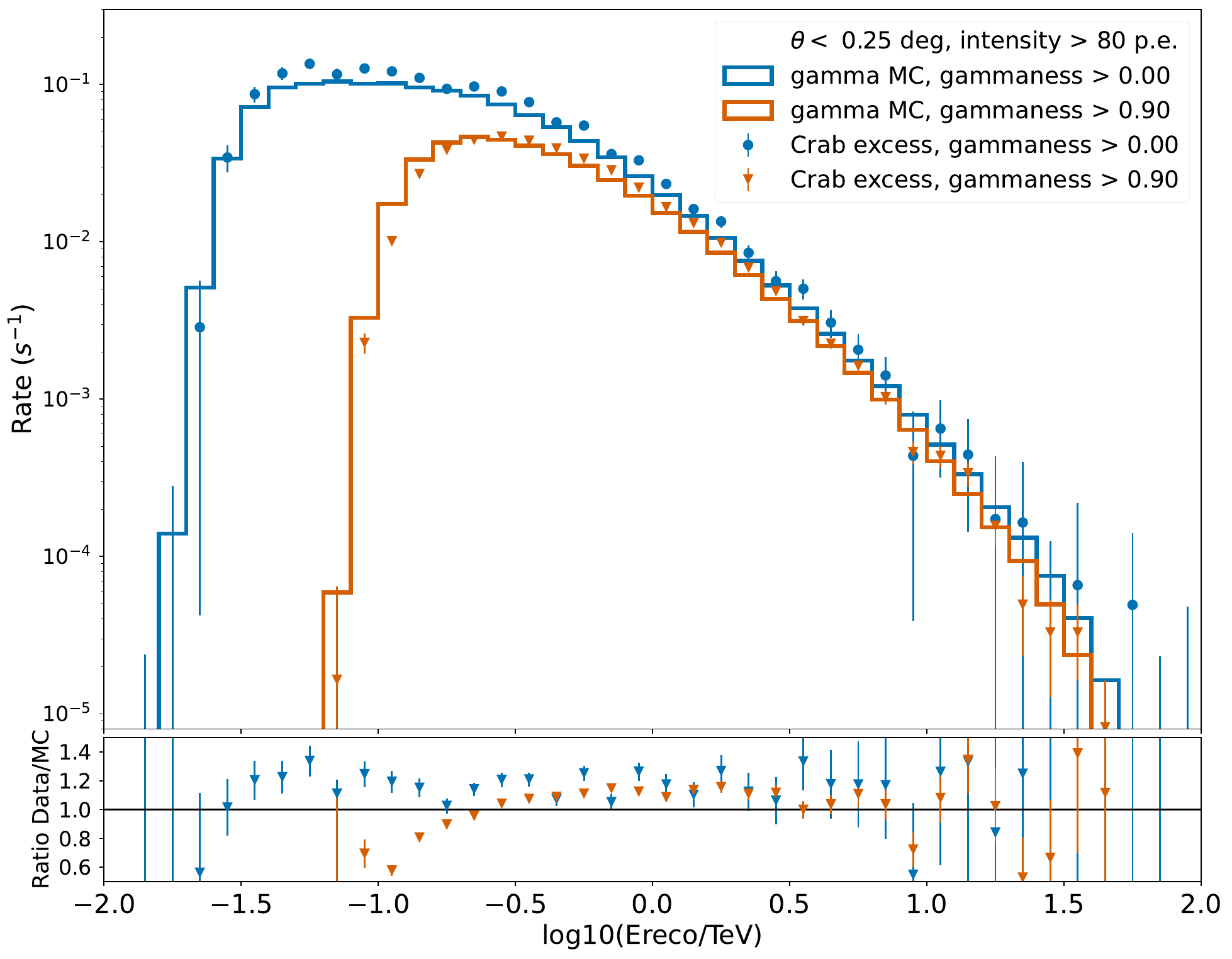}
    \caption{Comparison of the reconstructed photon energy distribution between data and MC.}
    \label{fig:hl_param_datamc_energy}
\end{figure}

A subset of parameter distribution comparison is shown below with both model parameters (Figs.~\ref{fig:model_param_datamc}-\ref{fig:model_param_datamc_length}) and primary particle parameters reconstructed by RFs (Figs.~\ref{fig:hl_param_datamc_gammaness}-\ref{fig:hl_param_datamc_energy}). 
In Figs.~\ref{fig:model_param_datamc}-\ref{fig:model_param_datamc_length}-\ref{fig:hl_param_datamc_gammaness}, the excess distribution in the data is shown as orange points. 
It is compared to the blue histogram obtained with the gamma-ray simulations. 
In histograms corresponding to the lowest intensity events, a pink step histogram represents the contamination of the OFF region by signal, which can occur because of the occasional poor direction reconstruction at low energies. A splitting of the data was performed depending on the intensity of the image. 
This allows us to see the evolution of the agreement with the image brightness. 
Faint images are harder to reconstruct due to a lower level of signal over baseline fluctuations in the waveform, fewer pixels containing a signal from which morphological information can be extracted, and a larger similarity between electromagnetic and hadronic showers.  
We can see in Fig.~\ref{fig:model_param_datamc} the good agreement between signal excess in the data and gamma-ray simulations for images with high intensity, and thus a good signal-to-noise ratio. The parameters shown are quite important for the reconstruction (see Fig.~\ref{fig:rf4-noclass}). 
When looking at the effect of image intensity on the agreement between data and MC, some problematic trends can be seen. 
For example, Fig.~\ref{fig:model_param_datamc_length} shows that the LH~fit length of images in high-intensity data is on average slightly larger than in simulations.

The effect of these small deviations between the observed and simulated distributions of the fit model parameters can be evaluated using the reconstructed particle properties. 
Figure~\ref{fig:hl_param_datamc_gammaness} shows the comparison for the gammaness for four image intensity ranges. Excellent agreement is found for images at low intensities but it degrades slowly at higher intensities. The distribution in the data is shifting slightly toward lower gammaness values. This indicates a lower gamma-hadron separation power in real data for these events,
but with a limited effect on the gamma-hadron separation power, since the score of hadrons is very low for images of this quality. A more problematic consequence is a wrong estimation of the effective area for a given event selection. With the $\theta < 0.25^\circ$ selection applied here, and assuming a selection of gammaness for a gamma-ray efficiency of 70\% per intensity bin, the true effective area would be biased compared to the expected one by, respectively, -4.6\%, +2.7\%, -8.7\%, and -16.9\%.  At very low intensity, a small excess of events with gammaness around 0.5 is seen.
The vicinity of the Crab Nebula is a rather complicated region for astrophysical observations. It is characterized by a high level of nonuniform NSB due to the presence of bright stars with a V-band magnitude below 7. This can lead to large statistical fluctuations in the levels of observed signal-like and background-like events, and to possible systematic bias in the inputs of the signal or background discriminator. In particular, the addition of light in pixels affected by stars can widen the light pool and create less elliptical images from EASs, which is thus more similar to hadron-initiated air showers. Given the high importance of extension parameters in the gamma-hadron classifier, this can naturally lead to a degradation of the classification power. But the full effect of stars is likely more complex, as it also biases the image intensity used to separate events in our figures, and very bright stars can also induce local reductions of the gain in the camera that are not accounted for in this analysis. 
Another possible source of discrepancy is the variation in trigger settings, which is pronounced in the early commissioning data of the LST-1, collected before September 2021. This was already discussed in~\citep{lstperf}, and no visible discrepancies arise from the variation in trigger settings when considering only events with an intensity above 80 p.e., so it should not affect our results.
Finally, the very good agreement for the distribution of the reconstructed energies is shown in Fig.~\ref{fig:hl_param_datamc_energy}.

\section{Performance with LST-1 simulations}
\label{sec:perf}

\begin{figure}[ht]
    \centering
    \includegraphics[width=\linewidth]{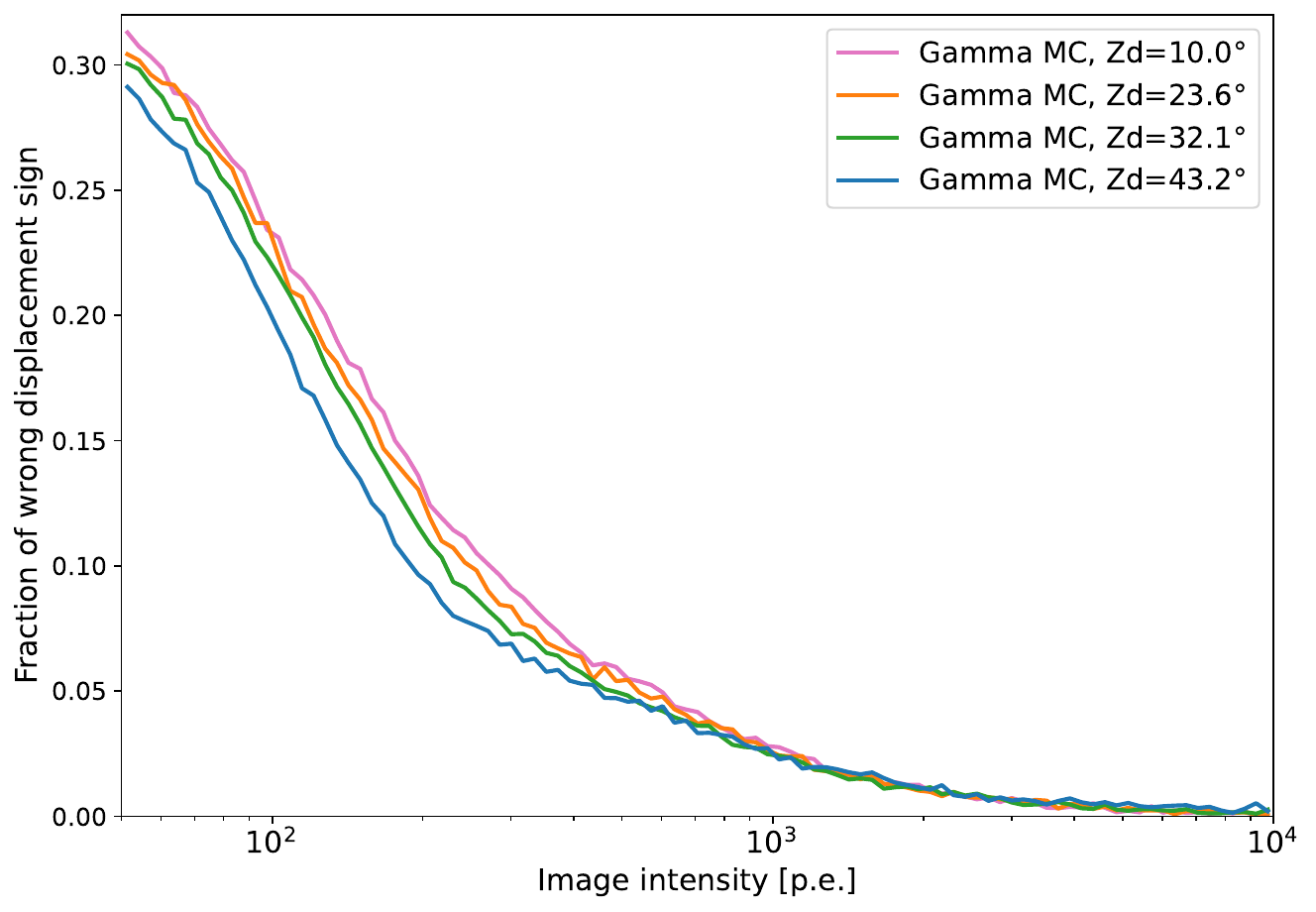}
    \caption{Fraction of gamma-ray events from our application MC reconstructed with a wrong sign from the displacement classifier as a function of image brightness after applying an energy-dependent gammaness cut for 70\% gamma-ray efficiency.}
    \label{fig:dispsignrate}
\end{figure}

To evaluate the performance improvement from our method, we extracted the angular resolution of the direction reconstruction as well as the relative resolution and bias of the reconstructed energy. 
We then compared it with the one used in the recent LST performance paper~\citep{lstperf} -- which we label “standard.” To ensure the fairness of the comparison, we reproduced the exact same event selection criteria and computation methods.
Since for low zenith angles, such as the ones considered here, the performance obtained with different azimuth values of the same elevation are nearly identical, we present average values over both azimuth values for each zenith. 
During direction reconstruction at low energy, the sign defining the orientation of the reconstructed vector can be wrong. The rate of such occurrences for gamma-ray MC as a function of image intensity is shown in Fig.~\ref{fig:dispsignrate}.
This appears as a secondary bump in the radial distribution of events. 
In order to keep an efficient angular event selection, and to only consider the central PSF for the angular resolution, both the $\theta$-based event selection and the angular resolution were evaluated only using events reconstructed with the right sign from the displacement classifier.

We applied the following event selection:
\begin{itemize}
    \item a reconstructed energy-dependent lower limit on the gammaness chosen to achieve a given gamma-ray efficiency (here 40, 70, or 90\%);
    \item for the angular resolution, a selection of events with a correct sign from the displacement classifier.
    \item for the effective area, energy resolution, and energy bias, a reconstructed energy-dependent cut on $\theta$ for a 70\% gamma efficiency evaluated on the gammaness selected events with a correct sign from the displacement classifier. The criteria on the sign from the displacement classifier was not directly applied in these cases.
\end{itemize} 
It is important to remember that the MCs used were uniformly tuned to the level of NSB corresponding to the Crab Nebula field of view. This field of view is in the galactic plane, and thus displays a higher NSB than that in the extragalactic sky. For both methods, slightly better results are expected if we consider observations with a lower NSB. The largest effect of NSB on our performance is a 5\% degradation of the angular resolution below 200 GeV compared to our nominal MC, with NSB levels slightly darker than a standard extragalactic field of view. Doubling the NSB injection degrades the angular resolution further by up to 10 percent in this energy range. Effects on the energy reconstruction are less than 5 percent in both cases and affect less of the energy range.

\begin{figure}[ht]
    \centering
    \includegraphics[width=\linewidth]{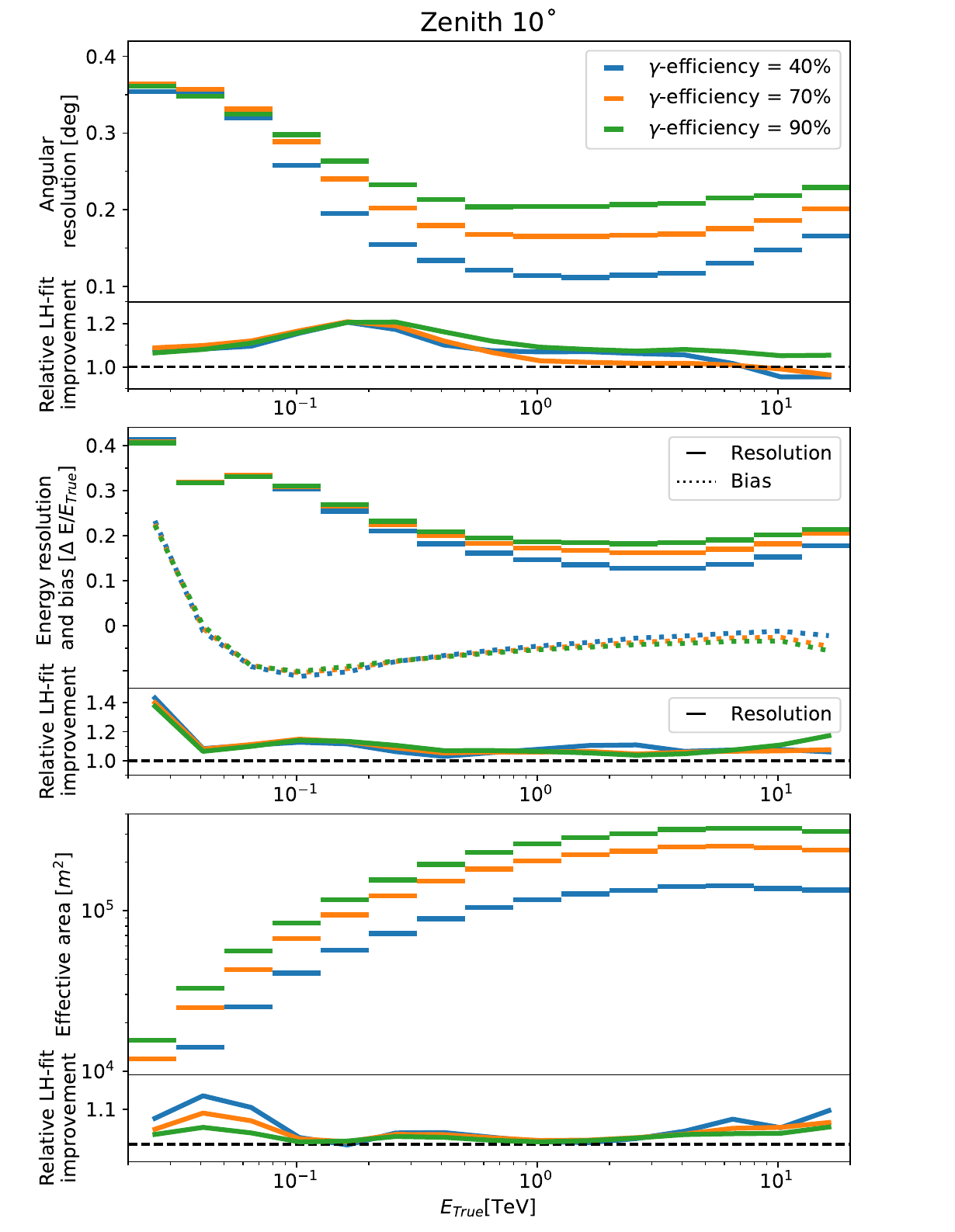}
    \caption{Performance of the likelihood reconstruction method at 10° from the zenith for three $\gamma$ efficiencies. Each plot shows the LH~fit performance in the top section and the relative improvement compared to the standard analysis, with performance evaluated in the exact same way, in the bottom section. \textit{Top:}~Angular resolution (68\% containment angle). \textit{Middle:}~Energy resolution (68\% relative containment) and bias (median shift). \textit{Bottom:}~Effective area. }
    \label{fig:perfzd10}
\end{figure}

In Fig.~\ref{fig:perfzd10}, the effect of the efficiency of the cut used to select events is evaluated for pointing at 10° away from the zenith. 
This allows us to see, without optimizing for a specific science case, the range of performances that could be reached depending on the requirement of event statistics versus reconstruction quality. 
The angular resolution, defined as the 68\% containment angle of the $\theta$ distribution of gamma-ray events, of LST is optimal in the tera-electronvolt energy region, where it achieves 0.11° considering the 40\% most gamma-like events and still reaches 0.20° if 90\% of the gamma-rays are retained. 
It degrades at low energy to 0.36° at 20~GeV. Such a degraded angular resolution can be problematic for the typical reflected background method used to analyze IACT data taken in “wobble mode,”\footnote{Wobble mode observations are performed by pointing the telescope at a position in the sky offset from the source of interest by a small angle (typically by 0.4° for LST), changing pointing regularly around the source position while keeping the same offset. 
Generally, pointings go in pairs, which are symmetric with respect to the source position. 
This allows us to estimate with the same dataset the background at the source position in a region of the sky with the same offset to the telescope pointing, and thus, assuming radial symmetry, with the same acceptance. Asymmetries potentially arising from the observation conditions are partially compensated for by the pointing pair and even more by using multiple such pairs.} since the region used to estimate the background is likely to be contaminated by the signal.
The LH~fit allows for an improvement of the angular resolution of 10 to 21\% at low energy, with a maximum improvement of around 150--200~GeV. The improvement in the full energy range is better when considering more events instead of only the most gamma-like ones, but an improvement is visible anyway. Indeed, the LH~fit angular resolution is $\sim$10\%  better than the standard analysis at nearly all energies. However, for the most gamma-like events, a worsening of a few percent is observed above $\sim$7~TeV. 
The energy resolution is also best near 2~TeV, reaching between 12.7 and 18.2\%. It is worse at 20~GeV, where it degrades to $\sim$40\%. 
The LH~fit allows for an improvement of the energy resolution of up to 43\% at threshold energy but is more generally around 10 to 15\% better than the standard analysis over the majority of the energy range considered, even at the highest energies. 
The difference between the effective areas is directly linked to the ratio of cut effectiveness. The effective areas reach a few $10^5 \text{ m}^2$ around a few hundred giga-electronvolts. The superior direction reconstruction of the events with LH~fit, coming from a better evaluation of the sign of the displacement vector, leads to an increase in the effective area at the lowest energies. At higher energy, the small differences in the effective area are linked to the different energy reconstructions with the two pipelines. The increase in effective area at high energy may be related to the degradation of angular resolution, since it implies the use of different events.
Improvements in the reconstruction quality at low energy are related to a few advantages of our method. 
First, no intensity-based cleaning was applied to select pixels, so the tails of the charge distribution -- which can be a non-negligible part of the signal at low energy -- were used with our method. 
Second, the timing of the signal is part of the fit. So, we constrained the shower direction with both time and geometric considerations, and we avoided using the charge information from a time in the waveform dominated by NSB, as it can occur during standard charge extraction in faint pixels.

Similar behaviors were observed with MC simulation with pointing at 23°, 32°, and 43° away from the zenith with a slight shift in energy. With these pointings, improvements compared to the standard pipeline are still mostly between 10\% and 20\% in angular and energy resolutions over most of the energy range. The maximum improvements are, respectively, 22\%, 22\%, and 22\% for the angular resolution and 47\%, 46\%, and 44\% for the energy resolution.
\newline

\noindent\textbf{\textit{Source-dependent analysis}}

\begin{figure}[ht]
    \centering
    \includegraphics[width=\linewidth]{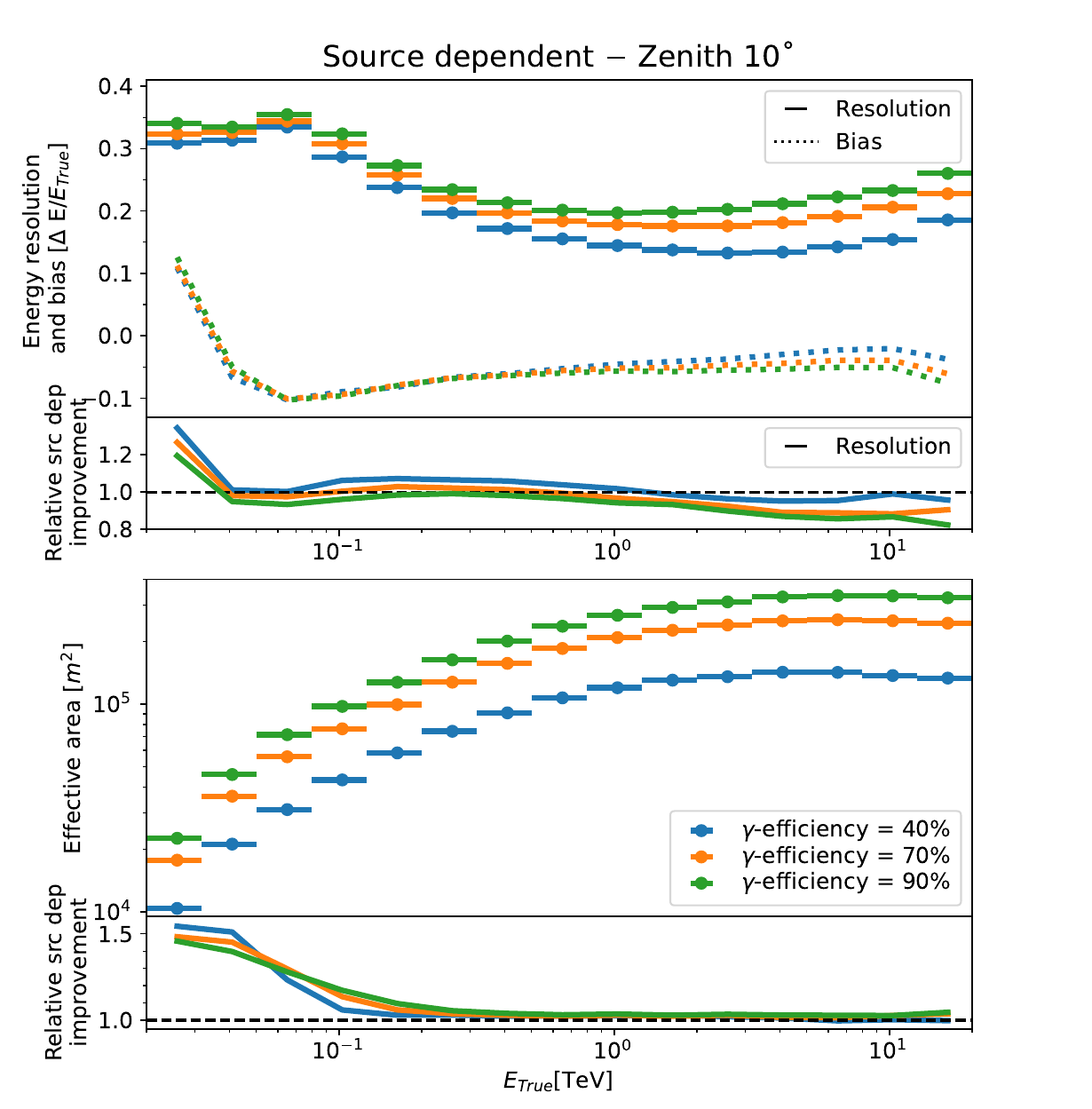}
    \caption{Performance of the likelihood reconstruction method at 10° from the zenith for three $\gamma$ efficiencies. Each plot shows the LH~fit source-dependent analysis performance in the top section and the relative improvement compared to the LH~fit source-independent analyses in the bottom section. \textit{Top:}~Energy resolution (68\% relative containment) and bias (median shift). \textit{Bottom:}~Effective area. }
    \label{fig:depperfzd10}
\end{figure}

With the source-dependent analysis, the position of the source in the camera is assumed to be known. In this case, a preselection based on $\theta$ used in the source-independent analysis cannot be used. Instead, we used a reconstructed energy-dependent cut on $\alpha$, the angle between the longer axis of the model and the line connecting the centroid of the model and the position of the source, for a 70\% gamma efficiency on the gammaness-selected events. Also, the preselection based on the sign from the displacement classifier was not carried out.
The latter leads to a better effective area at low energy in the event selection scheme used here. 
The performance of the LH~fit source-dependent analysis is shown in Fig.~\ref{fig:depperfzd10}. 
In this figure, the ratios indicate the improvement of the source-dependent analysis compared to the source-independent case both using the LH~fit method. 
An improvement of the energy resolution at the threshold is observed with up to 40\% improvement for the most gamma-like events with observations at 10° from the zenith. This is due to the fact that using the true source direction removes degeneracy in the implicit determination of the impact parameter, which is of high importance during the energy reconstruction.
Improvements of ~20\% are also observed for looser event selections. This is accompanied by and correlated with a large reduction in the energy bias. Over most of the energy range, the source-dependent and independent analyses show very similar results.

\section{Application to data: Crab Nebula analysis}
\label{sec:crab}

\begin{figure}[ht]
    \centering
    \includegraphics[width=\linewidth]{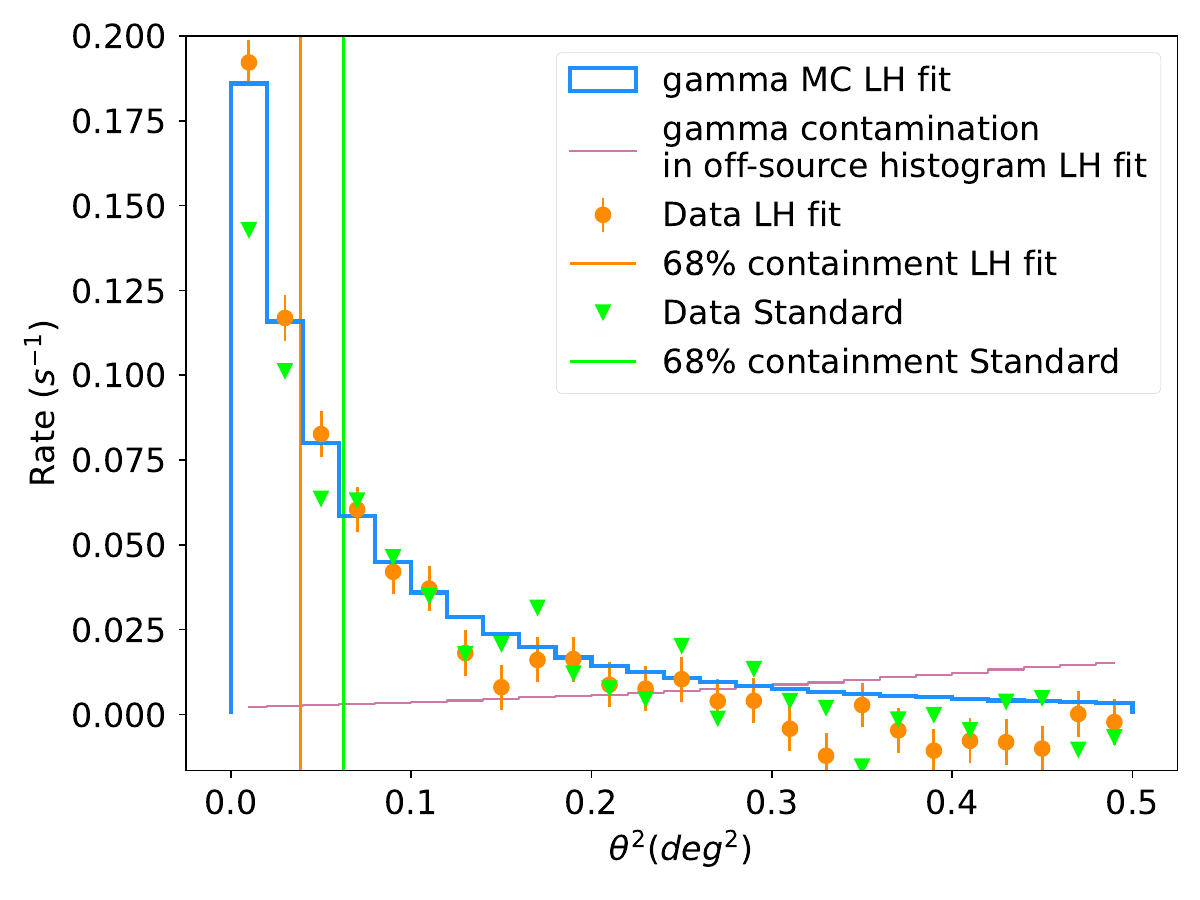}
    \caption{Distribution of the square angular distance between the source position and the reconstructed gamma-ray origin ($\theta^2$) for low intensity (80-200 p.e.) excess events. A good agreement is seen between data from Crab Nebula observations and expectations from MC simulation with the likelihood reconstruction. The same distribution for the standard reconstruction (from~\citep{lstperf}) is also displayed. Vertical lines represent the 68\% containment for both data distributions and show that the likelihood reconstruction reaches a better angular resolution.}
    \label{fig:thetalowint}
\end{figure}

Using the observations of the Crab Nebula, we performed three analyses. First, the improvement of the angular resolution seen on MC in the previous section was verified by comparing the distribution of theta for excess events in the case of low-intensity events, between the likelihood reconstruction and the standard one. Similarly to Sect.~\ref{sec:datamc}, Fig.~\ref{fig:thetalowint} shows a comparison of Crab Nebula data and an MC simulation, here for the square of the parameter $\theta$. It is here limited to the low image intensity case (80-200 p.e.) and also includes the same distribution for the standard reconstruction from~\citep{lstperf}. The comparison can be considered fair, since the gammaness cut applied for event selection is based on the same gamma-ray efficiency (80\%) for both. We considered the low image intensity case in order to verify the angular resolution improvement at low energy where it should be the largest. From the Crab data histograms, the 68\% angular containment was extracted: 0.196° for the likelihood reconstruction and 0.249° for the standard reconstruction. This corresponds to a 27\% improvement in the angular resolution, in line with the low energy estimate from simulations.

\begin{figure}[ht]
    \centering
    \includegraphics[width=\linewidth]{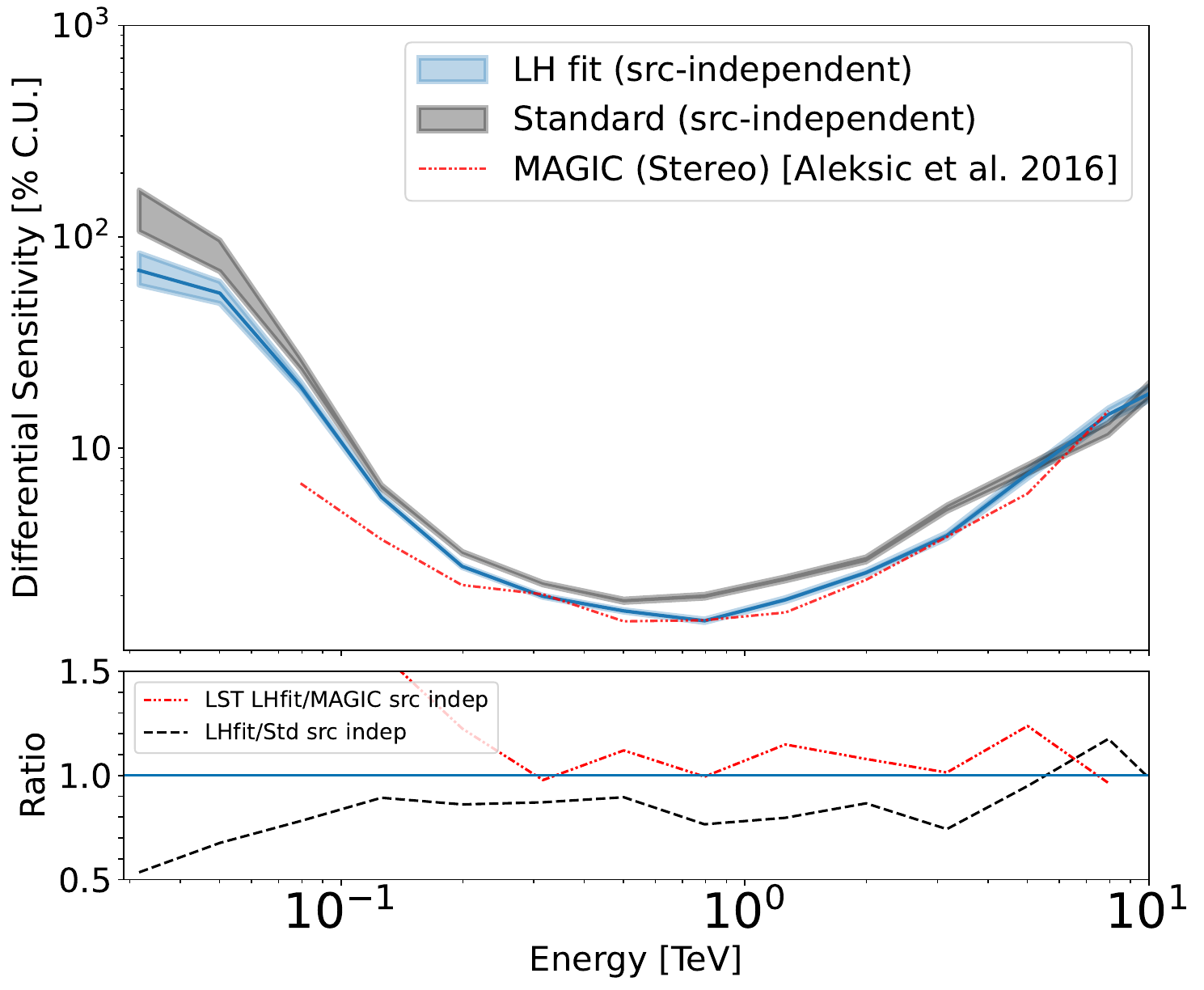}
    \caption{Differential sensitivity of LST-1 using the likelihood reconstruction method in percentage of the Crab Nebula flux. This was obtained from data by optimizing the gammaness and angular cuts for best sensitivity. The sensitivity shown here for the likelihood reconstruction, and associated statistical errors, are the average of the curves obtained through reversing the half of events used for cut optimization and sensitivity estimation. The “standard” sensitivity is from~\citep{lstperf}.}
    \label{fig:sensi}
\end{figure}

Second, we evaluated the detection potential of the analysis by evaluating the differential sensitivity\footnote{\label{note1}Defined as the minimal flux needed in an energy bin to reach a 5$\sigma$ detection with 50 h of observations while selecting at least 10 signal events with a signal/background of at least 5\% and with an acceptance ratio (source region/background only region) of 0.2.} from our dataset. 
To do so, an optimization of the gammaness and angular cuts was performed for each energy bin on half of the available events. The selection cuts thus optimized were applied to the other half. 
The sensitivity curve is shown in Fig.~\ref{fig:sensi}, where it is also compared to the standard analysis sensitivity obtained in the same way. An improvement is visible over the full energy range. Our method has a 10-20\% better flux sensitivity between 100~GeV and 5~TeV, nearly reaching the stereoscopic sensitivity of MAGIC above 300~GeV. The improvement increases rapidly below 100~GeV, to nearly a factor of two with respect to the standard analysis at 30~GeV. At these energies, the requirement of at least 5\% signal over the background ratio limits the sensitivity. The factor-two improvement needs to be considered carefully, since statistical and potential systematic errors can be large at the energy threshold. But the improvement trend below 100~GeV, associated with a better background rejection potential, should be real.

\begin{figure*}[ht]
    \centering
    \includegraphics[width=\textwidth]{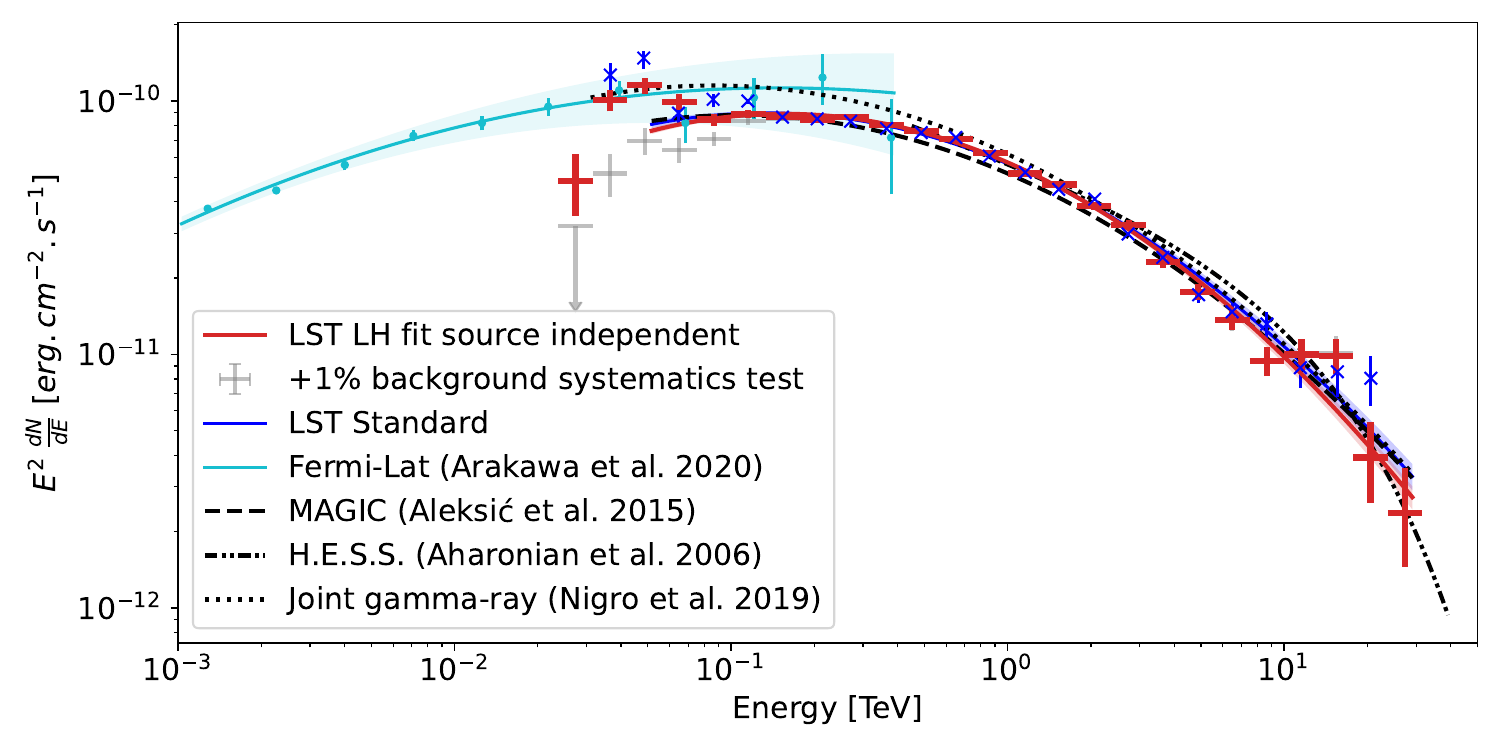}
    \caption{SED of the Crab Nebula obtained with the source-independent analysis presented in this paper and with the standard analysis from~\citep{lstperf}. The only errors are statistical.}
    \label{fig:sed}
\end{figure*}

\begin{figure*}[ht]
    \centering
    \includegraphics[width=\textwidth]{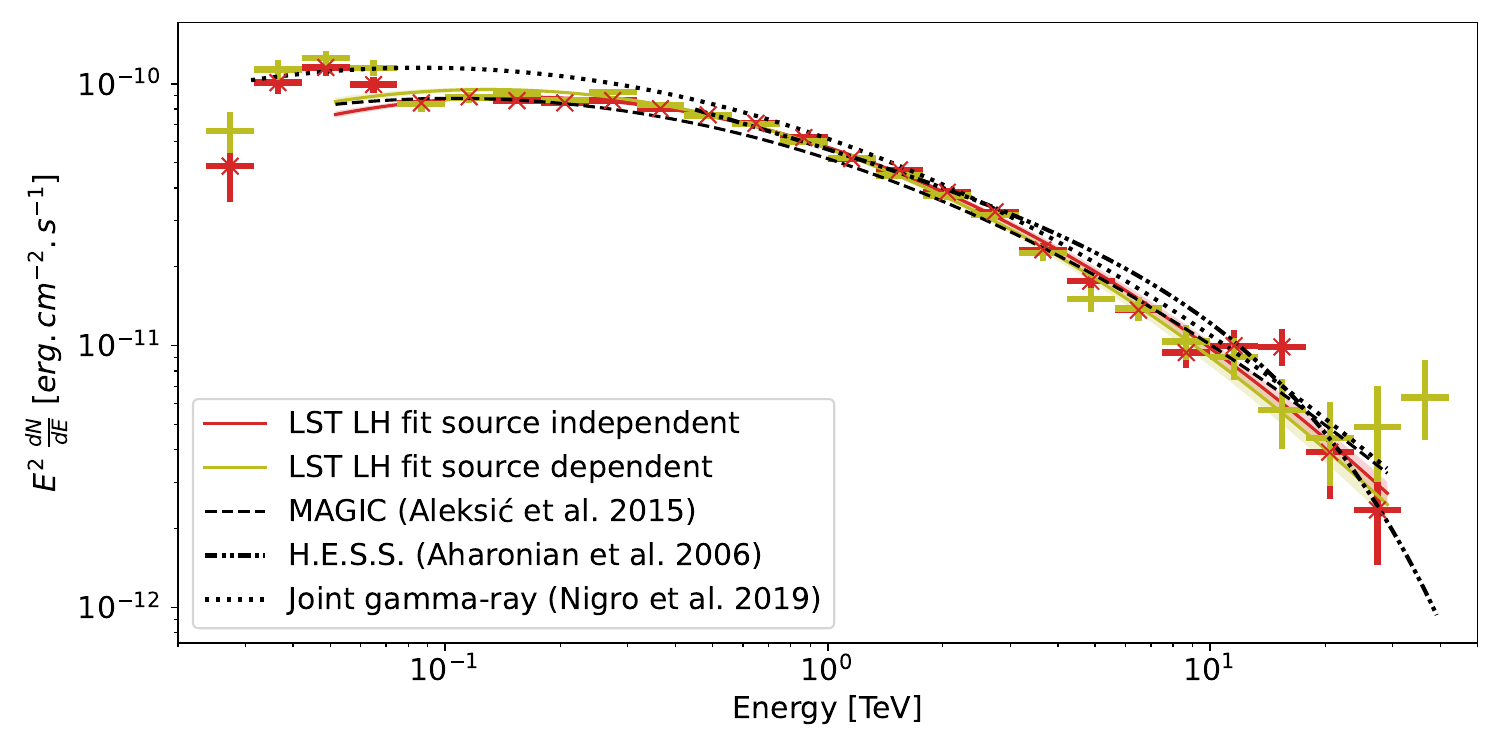}
    \caption{Comparison of the Crab Nebula SED obtained with the source-independent and source-dependent analysis using our reconstruction method.}
    \label{fig:depsed}
\end{figure*}

We also performed a high-level analysis using \textit{gammapy} to produce an SED (Fig.~\ref{fig:sed}). To do so, we applied event selection cuts derived from MC simulations following the procedure described in Sect.~\ref{sec:perf}, except that events with an intensity of less than 80 p.e. were removed. 
While the rejection of very faint, non-cosmic triggers and events too faint to be reconstructed correctly could still be achieved with an even lower threshold, the choice of 80 p.e. was motivated by the need to work around the evolving trigger settings used during the acquisition of this dataset. 
We performed the analysis using a 70\% efficiency gammaness cut and 70\% efficiency $\theta$ cut. The $\theta$ cut was in addition limited to 0.32° to allow for the use of the reflected background method.  
For each observation run, the closest MC simulation was used to derive the energy-dependent event selection cuts and produce IRFs. The event counts were evaluated in a region centered on the Crab position with an energy-dependent radius following the $\theta$ cut. The associated background count was evaluated using the reflected background method with one region taken symmetrically with respect to the center of the field of view. 
The spectral shape fit to the data is a log parabola function. A very good agreement is achieved with historical data from MAGIC~\citep{2015JHEAp...5...30A}, H.E.S.S.~\citep{2006A&A...457..899A}, and a joint (\textit{Fermi}-LAT, MAGIC, H.E.S.S. and Veritas) gamma-ray analysis~\citep{2019A&A...625A..10N}, while a signal is observed at energies lower than previous generation IACTs. 
The flux points were extracted after the SED using an energy binning of 8 bins per energy decade. At the lowest energies, there is a deviation between the fit spectral model and the flux points, which may be related to background systematics near the energy threshold of this dataset, as is investigated in~\citep{lstperf}, and with the computation of a flux point assuming a background count increased by 1\% in Fig~\ref{fig:sed}. The 1\% increase in the background count seems to overcorrect for the difference between the log-parabola spectrum and flux points. Thus, indicating that background count systematic errors should be lower than 1\%. Additionally, a smooth connection between LST observations at VHE and \textit{Fermi}-LAT observations at high energy~\citep{2020ApJ...897...33A} is observed. The source-dependent version of this SED is nearly identical, as is shown in Fig.\ref{fig:depsed}.

\section{Future potential}
\label{sec:future}

Although it is already possible to use the method presented in this paper with promising results, it can still be further improved. 
First, there could be an improvement in terms of processing time. The current version, for which extensive optimization work was done, processes events at a speed on the order of 15 events per second. 
Considering that the trigger rate of a single LST is between 5 and 10 kHz, a faster processing speed is desirable. A study of which events are the most time expensive, and of possible solutions, is thus interesting. One possible improvement could come from having a fast pre-analysis to remove very non-gamma-like events.
In addition, a higher level of optimization of the software, either through rewriting some sections or interfacing with a faster language, could lead to measurable improvements. 
Second, the current implementation does not make use of all the calibration information available, such as information on deactivated pixels and the temporal monitoring of pedestal variance from interleaved pedestal events. 
Including this information should improve performance when analyzing observation data and improve the agreement between observations and simulations.

The method implementation described in this paper was performed in a monoscopic context with LST-1. The extension of the technique to stereoscopic reconstruction is in preparation. 
It may require changing the model from a 2D image model, representing a Cherenkov shower projected in a camera plane, to a 3D shower model, representing the 3D distribution of Cherenkov light emitted by a photon-induced electromagnetic shower. 
The model would also need to be projected in all considered telescopes and the model parameters fit together. The alternative to applying the monoscopic parametrization to all telescopes, combining information at later stages, is also a possibility but would linearly scale the processing time with the number of telescopes.
Although the complexity per event will increase with a 3D model, both from the model and the quantity of data involved, it will bear advantages: the model will be closer to the primary particle and will thus directly include parameters that currently require RF to be recovered (in particular, the direction of arrival, but perhaps also the energy); the data available to constrain the model will increase faster than the model complexity. Three-dimensional shower models exist~\citep{2006APh....25..195L}, and would need to be improved and extended with a temporal component before implementation.

\section{Conclusion}
\label{sec:conclusion}
The likelihood-based method presented in this paper was successfully applied to the LST-1 data taken on the Crab Nebula and on gamma-ray simulations. In doing so, it was shown to be reliable for real applications, even with difficult fields of view. Our technique has been shown, from data or simulations, to improve the angular resolution by up to 22\%, the energy resolution by up to 47\%, and the sensitivity by a difference of up to nearly a factor of two at 30~GeV, compared to using Hillas parametrization with the same method to select events and derive these performance metrics. The greatest improvements are seen at low energies, where the biases linked to the charge extraction used in other methods are the largest. However, a general improvement over the full energy range is also observed, with both angular and energy resolution and sensitivity at least $\sim$10\% better at most energies. The improvements in angular and energy resolutions were verified to have limited dependence on the telescope pointing. Further developments and improvements of the method are envisioned. Computational optimization could increase the event processing speed. Exploiting the monitoring information during the observations could be included in the methods for better reconstruction. Finally, with the upcoming telescopes planned to be deployed in La Palma, the method can be adapted to stereoscopic reconstruction, potentially providing an improvement in performance in the CTAO era.

\begin{acknowledgements}
We gratefully acknowledge financial support from the following agencies and organisations: \newline
Conselho Nacional de Desenvolvimento Cient\'{\i}fico e Tecnol\'{o}gico (CNPq), Funda\c{c}\~{a}o de Amparo \`{a} Pesquisa do Estado do Rio de Janeiro (FAPERJ), Funda\c{c}\~{a}o de Amparo \`{a} Pesquisa do Estado de S\~{a}o Paulo (FAPESP), Funda\c{c}\~{a}o de Apoio \`{a} Ci\^encia, Tecnologia e Inova\c{c}\~{a}o do Paran\'a - Funda\c{c}\~{a}o Arauc\'aria, Ministry of Science, Technology, Innovations and Communications (MCTIC), Brasil;
Ministry of Education and Science, National RI Roadmap Project DO1-153/28.08.2018, Bulgaria;
Croatian Science Foundation, Rudjer Boskovic Institute, University of Osijek, University of Rijeka, University of Split, Faculty of Electrical Engineering, Mechanical Engineering and Naval Architecture, University of Zagreb, Faculty of Electrical Engineering and Computing, Croatia;
Ministry of Education, Youth and Sports, MEYS  LM2023047, EU/MEYS CZ.02.1.01/0.0/0.0/16\_013/0001403, CZ.02.1.01/0.0/0.0/18\_046/0016007, CZ.02.1.01/0.0/0.0/16\_019/0000754 , CZ.02.01.01/00/22\_008/0004632 and CZ.02.01.01/00/23\_015/0008197 Czech Republic;
CNRS-IN2P3, the French Programme d’investissements d’avenir and the Enigmass Labex, 
This work has been done thanks to the facilities offered by the Univ. Savoie Mont Blanc - CNRS/IN2P3 MUST computing center, France;
Max Planck Society, German Bundesministerium f{\"u}r Bildung und Forschung (Verbundforschung / ErUM), Deutsche Forschungsgemeinschaft (SFBs 876 and 1491), Germany;
Istituto Nazionale di Astrofisica (INAF), Istituto Nazionale di Fisica Nucleare (INFN), Italian Ministry for University and Research (MUR);
ICRR, University of Tokyo, JSPS, MEXT, Japan;
JST SPRING - JPMJSP2108;
Narodowe Centrum Nauki, grant number 2019/34/E/ST9/00224, Poland;
The Spanish groups acknowledge the Spanish Ministry of Science and Innovation and the Spanish Research State Agency (AEI) through the government budget lines PGE2021/28.06.000X.411.01, PGE2022/28.06.000X.411.01 and PGE2022/28.06.000X.711.04, and grants PID2022-139117NB-C44, PID2019-104114RB-C31,  PID2019-107847RB-C44, PID2019-104114RB-C32, PID2019-105510GB-C31, PID2019-104114RB-C33, PID2019-107847RB-C41, PID2019-107847RB-C43, PID2019-107847RB-C42, PID2019-107988GB-C22, PID2021-124581OB-I00, PID2021-125331NB-I00, PID2022-136828NB-C41, PID2022-137810NB-C22, PID2022-138172NB-C41, PID2022-138172NB-C42, PID2022-138172NB-C43, PID2022-139117NB-C41, PID2022-139117NB-C42, PID2022-139117NB-C43, PID2022-139117NB-C44, PID2022-136828NB-C42 funded by the Spanish MCIN/AEI/ 10.13039/501100011033 and “ERDF A way of making Europe; the ``Centro de Excelencia Severo Ochoa" program through grants no. CEX2019-000920-S, CEX2020-001007-S, CEX2021-001131-S; the ``Unidad de Excelencia Mar\'ia de Maeztu" program through grants no. CEX2019-000918-M, CEX2020-001058-M; the ``Ram\'on y Cajal" program through grants RYC2021-032991-I  funded by MICIN/AEI/10.13039/501100011033 and the European Union “NextGenerationEU”/PRTR; RYC2021-032552-I and RYC2020-028639-I; the ``Juan de la Cierva-Incorporaci\'on" program through grant no. IJC2019-040315-I and ``Juan de la Cierva-formaci\'on"' through grant JDC2022-049705-I. They also acknowledge the ``Atracción de Talento" program of Comunidad de Madrid through grant no. 2019-T2/TIC-12900; the project ``Tecnologi\'as avanzadas para la exploracio\'n del universo y sus componentes" (PR47/21 TAU), funded by Comunidad de Madrid, by the Recovery, Transformation and Resilience Plan from the Spanish State, and by NextGenerationEU from the European Union through the Recovery and Resilience Facility; the La Caixa Banking Foundation, grant no. LCF/BQ/PI21/11830030; Junta de Andaluc\'ia under Plan Complementario de I+D+I (Ref. AST22\_0001) and Plan Andaluz de Investigaci\'on, Desarrollo e Innovaci\'on as research group FQM-322; ``Programa Operativo de Crecimiento Inteligente" FEDER 2014-2020 (Ref.~ESFRI-2017-IAC-12), Ministerio de Ciencia e Innovaci\'on, 15\% co-financed by Consejer\'ia de Econom\'ia, Industria, Comercio y Conocimiento del Gobierno de Canarias; the ``CERCA" program and the grants 2021SGR00426 and 2021SGR00679, all funded by the Generalitat de Catalunya; and the European Union's ``Horizon 2020" GA:824064 and NextGenerationEU (PRTR-C17.I1). This research used the computing and storage resources provided by the Port d’Informació Científica (PIC) data center.
State Secretariat for Education, Research and Innovation (SERI) and Swiss National Science Foundation (SNSF), Switzerland;
The research leading to these results has received funding from the European Union's Seventh Framework Programme (FP7/2007-2013) under grant agreements No~262053 and No~317446;
This project is receiving funding from the European Union's Horizon 2020 research and innovation programs under agreement No~676134;
ESCAPE - The European Science Cluster of Astronomy \& Particle Physics ESFRI Research Infrastructures has received funding from the European Union’s Horizon 2020 research and innovation programme under Grant Agreement no. 824064.
\newline
This work was conducted in the context of the CTAO Consortium.
\end{acknowledgements}

%
\bibliographystyle{aa} 
\bibliography{biblio.bib} 

\begin{thebibliography}{33}
\expandafter\ifx\csname natexlab\endcsname\relax\def\natexlab#1{#1}\fi

\bibitem[{Abdalla {et~al.}(2020)}]{HESS:2019beq}
Abdalla, H. {et~al.} 2020, Nature Astron., 4, 167

\bibitem[{{Abe} {et~al.}(2023){Abe}, {Abe}, {Abe}, {Aguasca-Cabot}, {Agudo},
  {Alvarez Crespo}, {Antonelli}, {Aramo}, {Arbet-Engels}, {Arcaro}, {Artero},
  {Asano}, {Aubert}, {Baktash}, {Bamba}, {Baquero Larriva}, {Baroncelli},
  {Barres de Almeida}, {Barrio}, {Batkovic}, {Baxter}, {Becerra Gonz{\'a}lez},
  {Bernardini}, {Bernardos}, {Bernete Medrano}, {Berti}, {Bhattacharjee},
  {Biederbeck}, {Bigongiari}, {Bissaldi}, {Blanch}, {Bonnoli}, {Bordas},
  {Borghese}, {Bulgarelli}, {Burelli}, {Buscemi}, {Cardillo}, {Caroff},
  {Carosi}, {Cassol}, {Cauz}, {Ceribella}, {Chai}, {Cheng}, {Chiavassa},
  {Chikawa}, {Chytka}, {Cifuentes}, {Contreras}, {Cortina}, {Costantini},
  {D'Amico}, {Dalchenko}, {De Angelis}, {de Bony de Lavergne}, {De Lotto}, {de
  Menezes}, {Deleglise}, {Delgado}, {Delgado Mengual}, {della Volpe},
  {Dellaiera}, {Depaoli}, {Di Piano}, {Di Pierro}, {Di Tria}, {Di Venere},
  {D{\'\i}az}, {Dominik}, {Dominis Prester}, {Donini}, {Dorner}, {Doro},
  {Els{\"a}sser}, {Emery}, {Escudero}, {Fallah Ramazani}, {Ferrara},
  {Ferrarotto}, {Fiasson}, {Freixas Coromina}, {Fr{\"o}se}, {Fukami},
  {Fukazawa}, {Garcia}, {Garcia L{\'o}pez}, {Gasbarra}, {Gasparrini}, {Geyer},
  {Giesbrecht Paiva}, {Giglietto}, {Giordano}, {Giro}, {Gliwny}, {Godinovic},
  {Grau}, {Green}, {Green}, {Gunji}, {Hackfeld}, {Hadasch}, {Hahn},
  {Hashiyama}, {Hassan}, {Hayashi}, {Heckmann}, {Heller}, {Herrera Llorente},
  {Hirotani}, {Hoffmann}, {Horns}, {Houles}, {Hrabovsky}, {Hrupec}, {Hui},
  {H{\"u}tten}, {Iarlori}, {Imazawa}, {Inada}, {Inome}, {Ioka}, {Iori},
  {Ishio}, {Iwamura}, {Jacquemont}, {Jimenez Martinez}, {Jurysek}, {Kagaya},
  {Karas}, {Katagiri}, {Kataoka}, {Kerszberg}, {Kobayashi}, {Kong}, {Kubo},
  {Kushida}, {Lainez}, {Lamanna}, {Lamastra}, {Le Flour}, {Linhoff}, {Longo},
  {L{\'o}pez-Coto}, {L{\'o}pez-Moya}, {L{\'o}pez-Oramas}, {Loporchio},
  {Lorini}, {Luque-Escamilla}, {Majumdar}, {Makariev}, {Mandat}, {Manganaro},
  {Manic{\`o}}, {Mannheim}, {Mariotti}, {Marquez}, {Marsella}, {Mart{\'\i}},
  {Martinez}, {Mart{\'\i}nez}, {Mart{\'\i}nez}, {Marusevec}, {Mas-Aguilar},
  {Maurin}, {Mazin}, {Mestre Guillen}, {Micanovic}, {Miceli}, {Miener},
  {Miranda}, {Mirzoyan}, {Mizuno}, {Molero Gonzalez}, {Molina}, {Montaruli},
  {Monteiro}, {Moralejo}, {Morcuende}, {Morselli}, {Mrakovcic}, {Murase},
  {Nagai}, {Nagataki}, {Nakamori}, {Nickel}, {Nievas}, {Nishijima}, {Noda},
  {Nosek}, {Nozaki}, {Ohishi}, {Ohtani}, {Oka}, {Okazaki}, {Okumura}, {Orito},
  {Otero-Santos}, {Palatiello}, {Paneque}, {Pantaleo}, {Paoletti}, {Paredes},
  {Pech}, {Pecimotika}, {Peresano}, {P{\'e}rez}, {Pietropaolo}, {Pirola},
  {Plard}, {Podobnik}, {Poireau}, {Polo}, {Pons}, {Prandini}, {Prast},
  {Principe}, {Priyadarshi}, {Prouza}, {Rando}, {Rhode}, {Rib{\'o}}, {Rizi},
  {Rodriguez Fernandez}, {Ruiz}, {Saito}, {Sakurai}, {Sanchez},
  {{\v{S}}ari{\'c}}, {Sato}, {Saturni}, {Schleicher}, {Schmuckermaier},
  {Schubert}, {Schussler}, {Schweizer}, {Seglar Arroyo}, {Silvia}, {Sitarek},
  {Sliusar}, {Spolon}, {Stri{\v{s}}kovi{\'c}}, {Strzys}, {Suda}, {Sunada},
  {Tajima}, {Takahashi}, {Takahashi}, {Takata}, {Takeishi}, {Tam}, {Tanaka},
  {Tateishi}, {Tejedor}, {Temnikov}, {Terada}, {Terauchi}, {Terzic}, {Teshima},
  {Tluczykont}, {Tokanai}, {Torres}, {Travnicek}, {Truzzi}, {Tutone},
  {Uhlrich}, {Vacula}, {Vallania}, {van Scherpenberg}, {V{\'a}zquez Acosta},
  {Verguilov}, {Viale}, {Vigliano}, {Vigorito}, {Vitale}, {Voutsinas}, {Vovk},
  {Vuillaume}, {Walter}, {Will}, {Yamamoto}, {Yamazaki}, {Yoshida},
  {Yoshikoshi}, {Zywucka}, {Bernl{\"o}hr}, {Gueta}, {Kosack}, {Maier}, \&
  {Watson}}]{lstperf}
{Abe}, H., {Abe}, K., {Abe}, S., {et~al.} 2023, \apj, 956, 80

\bibitem[{Abeysekara {et~al.}(2017)}]{Abeysekara:2017mjj}
Abeysekara, A.~U. {et~al.} 2017, Astrophys. J., 843, 39

\bibitem[{Acero {et~al.}(2023)Acero, Aguasca-Cabot, Buchner, Carreto~Fidalgo,
  Chen, Chromey, Contreras~Gonzalez, de~Bony~de Lavergne, de~Miranda~Cardoso,
  Deil, Donath, Giunti, Hinton, Jouvin, Khélifi, King, Lefaucheur, Lenain,
  Linhoff, López-Coto, Mohrmann, Morcuende, Nakashima, Nigro, Olivera-Nieto,
  Owen, Panny, Papadopoulos~Orfanos, Paz~Arribas, Pintore, Poon, Remy, Ruiz,
  Siejkowski, Sinha, Sipőcz, Spir-Jacob, Terrier, Tibaldo, Unbehaun, van
  Eldik, Vuillaume, Weinstein, \& Wood}]{gammapy101}
Acero, F., Aguasca-Cabot, A., Buchner, J., {et~al.} 2023, Gammapy: Python
  toolbox for gamma-ray astronomy

\bibitem[{{Aharonian} {et~al.}(2006){Aharonian}, {Akhperjanian}, {Bazer-Bachi},
  {Beilicke}, {Benbow}, {Berge}, {Bernl{\"o}hr}, {Boisson}, {Bolz}, {Borrel},
  {Braun}, {Breitling}, {Brown}, {B{\"u}hler}, {B{\"u}sching}, {Carrigan},
  {Chadwick}, {Chounet}, {Cornils}, {Costamante}, {Degrange}, {Dickinson},
  {Djannati-Ata{\"\i}}, {O'C. Drury}, {Dubus}, {Egberts}, {Emmanoulopoulos},
  {Espigat}, {Feinstein}, {Ferrero}, {Fiasson}, {Fontaine}, {Funk}, {Funk},
  {Gallant}, {Giebels}, {Glicenstein}, {Goret}, {Hadjichristidis}, {Hauser},
  {Hauser}, {Heinzelmann}, {Henri}, {Hermann}, {Hinton}, {Hofmann}, {Holleran},
  {Horns}, {Jacholkowska}, {de Jager}, {Kh{\'e}lifi}, {Komin}, {Konopelko},
  {Kosack}, {Latham}, {Le Gallou}, {Lemi{\`e}re}, {Lemoine-Goumard}, {Lohse},
  {Martin}, {Martineau-Huynh}, {Marcowith}, {Masterson}, {McComb}, {de
  Naurois}, {Nedbal}, {Nolan}, {Noutsos}, {Orford}, {Osborne}, {Ouchrif},
  {Panter}, {Pelletier}, {Pita}, {P{\"u}hlhofer}, {Punch}, {Raubenheimer},
  {Raue}, {Rayner}, {Reimer}, {Reimer}, {Ripken}, {Rob}, {Rolland}, {Rowell},
  {Sahakian}, {Saug{\'e}}, {Schlenker}, {Schlickeiser}, {Schwanke}, {Sol},
  {Spangler}, {Spanier}, {Steenkamp}, {Stegmann}, {Superina}, {Tavernet},
  {Terrier}, {Th{\'e}oret}, {Tluczykont}, {van Eldik}, {Vasileiadis}, {Venter},
  {Vincent}, {V{\"o}lk}, {Wagner}, \& {Ward}}]{2006A&A...457..899A}
{Aharonian}, F., {Akhperjanian}, A.~G., {Bazer-Bachi}, A.~R., {et~al.} 2006,
  \aap, 457, 899

\bibitem[{{Aharonian, F.} {et~al.}(2024){Aharonian, F.}, {Ait Benkhali, F.},
  {Aschersleben, J.}, {Ashkar, H.}, {Backes, M.}, {Baktash, A.}, {Barbosa
  Martins, V.}, {Batzofin, R.}, {Becherini, Y.}, {Berge, D.}, {Bernlöhr, K.},
  {Bi, B.}, {Böttcher, M.}, {Boisson, C.}, {Bolmont, J.}, {de Bony de
  Lavergne, M.}, {Borowska, J.}, {Bradascio, F.}, {Breuhaus, M.}, {Brose, R.},
  {Brown, A.}, {Brun, F.}, {Bruno, B.}, {Bulik, T.}, {Burger-Scheidlin, C.},
  {Bylund, T.}, {Caroff, S.}, {Casanova, S.}, {Cecil, R.}, {Celic, J.},
  {Cerruti, M.}, {Chambery, P.}, {Chand, T.}, {Chandra, S.}, {Chen, A.},
  {Chibueze, J.}, {Chibueze, O.}, {Cotter, G.}, {Cristofari, P.}, {Devin, J.},
  {Djannati-Ataï, A.}, {Djuvsland, J.}, {Dmytriiev, A.}, {Einecke, S.},
  {Ernenwein, J.-P.}, {Fegan, S.}, {Feijen, K.}, {Filipović, M.}, {Fontaine,
  G.}, {Füßling, M.}, {Funk, S.}, {Gabici, S.}, {Gallant, Y. A.}, {Giavitto,
  G.}, {Glawion, D.}, {Glicenstein, J. F.}, {Glombitza, J.}, {Goswami, P.},
  {Grolleron, G.}, {Grondin, M.-H.}, {Haerer, L.}, {Hinton, J. A.}, {Hofmann,
  W.}, {Holch, T. L.}, {Holler, M.}, {Horns, D.}, {Jamrozy, M.}, {Jankowsky,
  F.}, {Joshi, V.}, {Kasai, E.}, {Katarzyński, K.}, {Khatoon, R.}, {Khélifi,
  B.}, {Kluźniak, W.}, {Komin, Nu.}, {Kosack, K.}, {Kostunin, D.}, {Kundu,
  A.}, {Lang, R. G.}, {Le Stum, S.}, {Leitl, F.}, {Lemière, A.},
  {Lemoine-Goumard, M.}, {Lenain, J.-P.}, {Leuschner, F.}, {Luashvili, A.},
  {Mackey, J.}, {Malyshev, D.}, {Malyshev, D.}, {Marandon, V.}, {Marinos, P.},
  {Martí-Devesa, G.}, {Marx, R.}, {Mehta, A.}, {Meyer, M.}, {Mitchell, A.},
  {Moderski, R.}, {Mohrmann, L.}, {Montanari, A.}, {Moulin, E.}, {Murach, T.},
  {de Naurois, M.}, {Niemiec, J.}, {O’Brien, P.}, {Ohm, S.}, {Olivera-Nieto,
  L.}, {de Ona Wilhelmi, E.}, {Ostrowski, M.}, {Panny, S.}, {Panter, M.},
  {Parsons, R. D.}, {Peron, G.}, {Prokhorov, D. A.}, {Pühlhofer, G.}, {Punch,
  M.}, {Quirrenbach, A.}, {Regeard, M.}, {Reichherzer, P.}, {Reimer, A.},
  {Reimer, O.}, {Ren, H.}, {Renaud, M.}, {Reville, B.}, {Rieger, F.},
  {Roellinghoff, G.}, {Rudak, B.}, {Sahakian, V.}, {Salzmann, H.}, {Sasaki,
  M.}, {Schüssler, F.}, {Schutte, H. M.}, {Shapopi, J. N. S.}, {Specovius,
  A.}, {Spencer, S.}, {Stawarz, Ł.}, {Steenkamp, R.}, {Steinmassl, S.},
  {Steppa, C.}, {Streil, K.}, {Sushch, I.}, {Suzuki, H.}, {Takahashi, T.},
  {Tanaka, T.}, {Terrier, R.}, {Tluczykont, M.}, {Tsuji, N.}, {Unbehaun, T.},
  {van Eldik, C.}, {Vecchi, M.}, {Veh, J.}, {Venter, C.}, {Vink, J.}, {Wach,
  T.}, {Wagner, S. J.}, {Wierzcholska, A.}, {Zacharias, M.}, {Zargaryan, D.},
  {Zdziarski, A. A.}, {Zech, A.}, {Zouari, S.}, {Żywucka, N.}, \& {Harding,
  A.}}]{Crab_FermiHESS2024}
{Aharonian, F.}, {Ait Benkhali, F.}, {Aschersleben, J.}, {et~al.} 2024, \aap,
  686, A308

\bibitem[{{Aleksi{\'c}} {et~al.}(2015){Aleksi{\'c}}, {Ansoldi}, {Antonelli},
  {Antoranz}, {Babic}, {Bangale}, {Barrio}, {Becerra Gonz{\'a}lez}, {Bednarek},
  {Bernardini}, {Biasuzzi}, {Biland}, {Blanch}, {Bonnefoy}, {Bonnoli},
  {Borracci}, {Bretz}, {Carmona}, {Carosi}, {Colin}, {Colombo}, {Contreras},
  {Cortina}, {Covino}, {Da Vela}, {Dazzi}, {De Angelis}, {De Caneva}, {De
  Lotto}, {de O{\~n}a Wilhelmi}, {Delgado Mendez}, {Doert}, {Dominis Prester},
  {Dorner}, {Doro}, {Einecke}, {Eisenacher}, {Elsaesser}, {Fonseca}, {Font},
  {Frantzen}, {Fruck}, {Galindo}, {Garc{\'\i}a L{\'o}pez}, {Garczarczyk},
  {Garrido Terrats}, {Gaug}, {Godinovi{\'c}}, {Gonz{\'a}lez Mu{\~n}oz},
  {Gozzini}, {Hadasch}, {Hanabata}, {Hayashida}, {Herrera}, {Hildebrand},
  {Hose}, {Hrupec}, {Idec}, {Kadenius}, {Kellermann}, {Kodani}, {Konno},
  {Krause}, {Kubo}, {Kushida}, {La Barbera}, {Lelas}, {Lewandowska},
  {Lindfors}, {Lombardi}, {L{\'o}pez}, {L{\'o}pez-Coto}, {L{\'o}pez-Oramas},
  {Lorenz}, {Lozano}, {Makariev}, {Mallot}, {Maneva}, {Mankuzhiyil},
  {Mannheim}, {Maraschi}, {Marcote}, {Mariotti}, {Mart{\'\i}nez}, {Mazin},
  {Menzel}, {Miranda}, {Mirzoyan}, {Moralejo}, {Munar-Adrover}, {Nakajima},
  {Niedzwiecki}, {Nilsson}, {Nishijima}, {Noda}, {Nowak}, {Orito},
  {Overkemping}, {Paiano}, {Palatiello}, {Paneque}, {Paoletti}, {Paredes},
  {Paredes-Fortuny}, {Persic}, {Prada Moroni}, {Prandini}, {Preziuso},
  {Puljak}, {Reinthal}, {Rhode}, {Rib{\'o}}, {Rico}, {Rodriguez Garcia},
  {R{\"u}gamer}, {Saggion}, {Saito}, {Saito}, {Satalecka}, {Scalzotto},
  {Scapin}, {Schultz}, {Schweizer}, {Shore}, {Sillanp{\"a}{\"a}}, {Sitarek},
  {Snidaric}, {Sobczynska}, {Spanier}, {Stamatescu}, {Stamerra}, {Steinbring},
  {Storz}, {Strzys}, {Takalo}, {Takami}, {Tavecchio}, {Temnikov}, {Terzi{\'c}},
  {Tescaro}, {Teshima}, {Thaele}, {Tibolla}, {Torres}, {Toyama}, {Treves},
  {Uellenbeck}, {Vogler}, {Wagner}, {Zanin}, {Horns}, {Mart{\'\i}n}, \&
  {Meyer}}]{2015JHEAp...5...30A}
{Aleksi{\'c}}, J., {Ansoldi}, S., {Antonelli}, L.~A., {et~al.} 2015, Journal of
  High Energy Astrophysics, 5, 30

\bibitem[{Aleksi\'c {et~al.}(2015)}]{MAGIC:2014izm}
Aleksi\'c, J. {et~al.} 2015, JHEA, 5-6, 30

\bibitem[{Alispach {et~al.}(2020)}]{Alispach:2020rqn}
Alispach, C. {et~al.} 2020, JINST, 15, P11010

\bibitem[{Alispach(2020)}]{cyrilthesis}
Alispach, C.~M. 2020, PhD thesis, University of Geneva, Switzerland, iD:
  unige:147894

\bibitem[{{Aliu} {et~al.}(2009){Aliu}, {Anderhub}, {Antonelli}, {Antoranz},
  {Backes}, {Baixeras}, {Barrio}, {Bartko}, {Bastieri}, {Becker}, {Bednarek},
  {Berger}, {Bernardini}, {Biland}, {Bock}, {Bonnoli}, {Bordas}, {Borla
  Tridon}, {Bosch-Ramon}, {Bretz}, {Britvitch}, {Camara}, {Carmona},
  {Chilingarian}, {Commichau}, {Contreras}, {Cortina}, {Costado}, {Covino},
  {Curtef}, {Dazzi}, {de Angelis}, {de Cea Del Pozo}, {de Los Reyes}, {de
  Lotto}, {de Maria}, {de Sabata}, {Delgado Mendez}, {Dominguez}, {Dorner},
  {Doro}, {Els{\"a}sser}, {Errando}, {Fagiolini}, {Ferenc}, {Fern{\'a}ndez},
  {Firpo}, {Fonseca}, {Font}, {Galante}, {Garc{\'\i}a L{\'o}pez},
  {Garczarczyk}, {Gaug}, {Goebel}, {Hadasch}, {Hayashida}, {Herrero},
  {H{\"o}hne}, {Hose}, {Hsu}, {Huber}, {Jogler}, {Kranich}, {La Barbera},
  {Laille}, {Leonardo}, {Lindfors}, {Lombardi}, {Longo}, {L{\'o}pez}, {Lorenz},
  {Majumdar}, {Maneva}, {Mankuzhiyil}, {Mannheim}, {Maraschi}, {Mariotti},
  {Mart{\'\i}nez}, {Mazin}, {Meucci}, {Meyer}, {Miranda}, {Mirzoyan}, {Moles},
  {Moralejo}, {Nieto}, {Nilsson}, {Ninkovic}, {Otte}, {Oya}, {Paoletti},
  {Paredes}, {Pasanen}, {Pascoli}, {Pauss}, {Pegna}, {Perez-Torres}, {Persic},
  {Peruzzo}, {Piccioli}, {Prada}, {Prandini}, {Puchades}, {Raymers}, {Rhode},
  {Rib{\'o}}, {Rico}, {Rissi}, {Robert}, {R{\"u}gamer}, {Saggion}, {Saito},
  {Salvati}, {Sanchez-Conde}, {Sartori}, {Satalecka}, {Scalzotto}, {Scapin},
  {Schweizer}, {Shayduk}, {Shinozaki}, {Shore}, {Sidro}, {Sierpowska-Bartosik},
  {Sillanp{\"a}{\"a}}, {Sitarek}, {Sobczynska}, {Spanier}, {Stamerra}, {Stark},
  {Takalo}, {Tavecchio}, {Temnikov}, {Tescaro}, {Teshima}, {Tluczykont},
  {Torres}, {Turini}, {Vankov}, {Venturini}, {Vitale}, {Wagner}, {Wittek},
  {Zabalza}, {Zandanel}, {Zanin}, \& {Zapatero}}]{2009APh....30..293A}
{Aliu}, E., {Anderhub}, H., {Antonelli}, L.~A., {et~al.} 2009, Astroparticle
  Physics, 30, 293

\bibitem[{Amato \& Olmi(2021)}]{Amato:2021gzt}
Amato, E. \& Olmi, B. 2021, Universe, 7, 448

\bibitem[{{Arakawa} {et~al.}(2020){Arakawa}, {Hayashida}, {Khangulyan}, \&
  {Uchiyama}}]{2020ApJ...897...33A}
{Arakawa}, M., {Hayashida}, M., {Khangulyan}, D., \& {Uchiyama}, Y. 2020, \apj,
  897, 33

\bibitem[{{Atwood} {et~al.}(2009){Atwood}, {Abdo}, {Ackermann}, {Althouse},
  {Anderson}, {Axelsson}, {Baldini}, {Ballet}, {Band}, {Barbiellini},
  {Bartelt}, {Bastieri}, {Baughman}, {Bechtol}, {B{\'e}d{\'e}r{\`e}de},
  {Bellardi}, {Bellazzini}, {Berenji}, {Bignami}, {Bisello}, {Bissaldi},
  {Blandford}, {Bloom}, {Bogart}, {Bonamente}, {Bonnell}, {Borgland},
  {Bouvier}, {Bregeon}, {Brez}, {Brigida}, {Bruel}, {Burnett}, {Busetto},
  {Caliandro}, {Cameron}, {Caraveo}, {Carius}, {Carlson}, {Casandjian},
  {Cavazzuti}, {Ceccanti}, {Cecchi}, {Charles}, {Chekhtman}, {Cheung},
  {Chiang}, {Chipaux}, {Cillis}, {Ciprini}, {Claus}, {Cohen-Tanugi},
  {Condamoor}, {Conrad}, {Corbet}, {Corucci}, {Costamante}, {Cutini}, {Davis},
  {Decotigny}, {DeKlotz}, {Dermer}, {de Angelis}, {Digel}, {do Couto e Silva},
  {Drell}, {Dubois}, {Dumora}, {Edmonds}, {Fabiani}, {Farnier}, {Favuzzi},
  {Flath}, {Fleury}, {Focke}, {Funk}, {Fusco}, {Gargano}, {Gasparrini},
  {Gehrels}, {Gentit}, {Germani}, {Giebels}, {Giglietto}, {Giommi}, {Giordano},
  {Glanzman}, {Godfrey}, {Grenier}, {Grondin}, {Grove}, {Guillemot}, {Guiriec},
  {Haller}, {Harding}, {Hart}, {Hays}, {Healey}, {Hirayama}, {Hjalmarsdotter},
  {Horn}, {Hughes}, {J{\'o}hannesson}, {Johansson}, {Johnson}, {Johnson},
  {Johnson}, {Johnson}, {Kamae}, {Katagiri}, {Kataoka}, {Kavelaars}, {Kawai},
  {Kelly}, {Kerr}, {Klamra}, {Kn{\"o}dlseder}, {Kocian}, {Komin}, {Kuehn},
  {Kuss}, {Landriu}, {Latronico}, {Lee}, {Lee}, {Lemoine-Goumard}, {Lionetto},
  {Longo}, {Loparco}, {Lott}, {Lovellette}, {Lubrano}, {Madejski}, {Makeev},
  {Marangelli}, {Massai}, {Mazziotta}, {McEnery}, {Menon}, {Meurer},
  {Michelson}, {Minuti}, {Mirizzi}, {Mitthumsiri}, {Mizuno}, {Moiseev},
  {Monte}, {Monzani}, {Moretti}, {Morselli}, {Moskalenko}, {Murgia},
  {Nakamori}, {Nishino}, {Nolan}, {Norris}, {Nuss}, {Ohno}, {Ohsugi}, {Omodei},
  {Orlando}, {Ormes}, {Paccagnella}, {Paneque}, {Panetta}, {Parent}, {Pearce},
  {Pepe}, {Perazzo}, {Pesce-Rollins}, {Picozza}, {Pieri}, {Pinchera}, {Piron},
  {Porter}, {Poupard}, {Rain{\`o}}, {Rando}, {Rapposelli}, {Razzano}, {Reimer},
  {Reimer}, {Reposeur}, {Reyes}, {Ritz}, {Rochester}, {Rodriguez}, {Romani},
  {Roth}, {Russell}, {Ryde}, {Sabatini}, {Sadrozinski}, {Sanchez}, {Sander},
  {Sapozhnikov}, {Parkinson}, {Scargle}, {Schalk}, {Scolieri}, {Sgr{\`o}},
  {Share}, {Shaw}, {Shimokawabe}, {Shrader}, {Sierpowska-Bartosik}, {Siskind},
  {Smith}, {Smith}, {Spandre}, {Spinelli}, {Starck}, {Stephens}, {Strickman},
  {Strong}, {Suson}, {Tajima}, {Takahashi}, {Takahashi}, {Tanaka}, {Tenze},
  {Tether}, {Thayer}, {Thayer}, {Thompson}, {Tibaldo}, {Tibolla}, {Torres},
  {Tosti}, {Tramacere}, {Turri}, {Usher}, {Vilchez}, {Vitale}, {Wang},
  {Watters}, {Winer}, {Wood}, {Ylinen}, \& {Ziegler}}]{fermilat}
{Atwood}, W.~B., {Abdo}, A.~A., {Ackermann}, M., {et~al.} 2009, \apj, 697, 1071

\bibitem[{Cao {et~al.}(2021)}]{LHAASO:2021cbz}
Cao, Z. {et~al.} 2021, Science, 373, 425

\bibitem[{ctapipe(2022)}]{local_peak_window_sum}
ctapipe. 2022, Local Peak Window Sum Algorithm,
  \url{https://cta-observatory.github.io/ctapipe/api/ctapipe.image.extractor.LocalPeakWindowSum.html#ctapipe.image.extractor.LocalPeakWindowSum}

\bibitem[{De~Angelis \& Mallamaci(2018)}]{DeAngelis:2018lra}
De~Angelis, A. \& Mallamaci, M. 2018, Eur. Phys. J. Plus, 133, 324

\bibitem[{{de Naurois} \& {Rolland}(2009)}]{2009APh....32..231D}
{de Naurois}, M. \& {Rolland}, L. 2009, Astroparticle Physics, 32, 231

\bibitem[{Dembinski \& et~al.(2020)}]{iminuit}
Dembinski, H. \& et~al., P.~O. 2020

\bibitem[{{Donath} {et~al.}(2023){Donath}, {Terrier}, {Remy}, {Sinha}, {Nigro},
  {Pintore}, {Kh{\'e}lifi}, {Olivera-Nieto}, {Ruiz}, {Br{\"u}gge}, {Linhoff},
  {Contreras}, {Acero}, {Aguasca-Cabot}, {Berge}, {Bhattacharjee}, {Buchner},
  {Boisson}, {Carreto Fidalgo}, {Chen}, {de Bony de Lavergne}, {de Miranda
  Cardoso}, {Deil}, {F{\"u}{\ss}ling}, {Funk}, {Giunti}, {Hinton}, {Jouvin},
  {King}, {Lefaucheur}, {Lemoine-Goumard}, {Lenain}, {L{\'o}pez-Coto},
  {Mohrmann}, {Morcuende}, {Panny}, {Regeard}, {Saha}, {Siejkowski},
  {Siemiginowska}, {Sip{\H{o}}cz}, {Unbehaun}, {van Eldik}, {Vuillaume}, \&
  {Zanin}}]{2023A&A...678A.157D}
{Donath}, A., {Terrier}, R., {Remy}, Q., {et~al.} 2023, \aap, 678, A157

\bibitem[{Fegan(1997)}]{Fegan:1997db}
Fegan, D.~J. 1997, J. Phys. G, 23, 1013

\bibitem[{Hillas(1985)}]{hillas_parameters}
Hillas, A.~M. 1985, in 19th International Cosmic Ray Conference (ICRC19),
  Vol.~3, 445

\bibitem[{Jacquemont {et~al.}(2019)Jacquemont, Vuillaume, Benoit, Maurin,
  Lambert, Lamanna, \& Brill}]{Jacquemont:2019pw}
Jacquemont, M., Vuillaume, T., Benoit, A., {et~al.} 2019, PoS, ICRC2019, 705

\bibitem[{{Lemoine-Goumard} {et~al.}(2006){Lemoine-Goumard}, {Degrange}, \&
  {Tluczykont}}]{2006APh....25..195L}
{Lemoine-Goumard}, M., {Degrange}, B., \& {Tluczykont}, M. 2006, Astroparticle
  Physics, 25, 195

\bibitem[{Lopez-Coto {et~al.}(2023)Lopez-Coto, Vuillaume, Moralejo, Linhoff,
  Cassol, Priyadarshi, Morcuende, Nozaki, Bernardos, Gliwny, Ruiz, Dalchenko,
  yrenier, Saha, Nickel, Sitarek, Alispach, Pillera, aaguasca, Andres-Baquero,
  Láinez, Takahashi, sn621, \& yiwamura}]{lstchain_zen}
Lopez-Coto, R., Vuillaume, T., Moralejo, A., {et~al.} 2023,
  cta-observatory/cta-lstchain: v0.9.14 – 2023-09-25

\bibitem[{{Mazin} {et~al.}(2008){Mazin}, {Bigongiari}, {Goebel}, {Moralejo}, \&
  {Wittek}}]{2008ICRC....5.1253M}
{Mazin}, D., {Bigongiari}, C., {Goebel}, F., {Moralejo}, A., \& {Wittek}, W.
  2008, in International Cosmic Ray Conference, Vol.~5, International Cosmic
  Ray Conference, 1253--1256

\bibitem[{Meagher(2016)}]{Meagher:2015igh}
Meagher, K. 2016, PoS, ICRC2015, 792

\bibitem[{{Miener} {et~al.}(2022){Miener}, {Nieto}, {Brill}, {Spencer}, \&
  {Contreras}}]{2022icrc.confE.730M}
{Miener}, T., {Nieto}, D., {Brill}, A., {Spencer}, S.~T., \& {Contreras}, J.~L.
  2022, in 37th International Cosmic Ray Conference, 730

\bibitem[{{Nigro} {et~al.}(2019){Nigro}, {Deil}, {Zanin}, {Hassan}, {King},
  {Ruiz}, {Saha}, {Terrier}, {Br{\"u}gge}, {N{\"o}the}, {Bird}, {Lin},
  {Aleksi{\'c}}, {Boisson}, {Contreras}, {Donath}, {Jouvin}, {Kelley-Hoskins},
  {Khelifi}, {Kosack}, {Rico}, \& {Sinha}}]{2019A&A...625A..10N}
{Nigro}, C., {Deil}, C., {Zanin}, R., {et~al.} 2019, \aap, 625, A10

\bibitem[{{Parsons} \& {Hinton}(2014)}]{2014APh....56...26P}
{Parsons}, R.~D. \& {Hinton}, J.~A. 2014, Astroparticle Physics, 56, 26

\bibitem[{{Spencer} {et~al.}(2021){Spencer}, {Armstrong}, {Watson}, {Mangano},
  {Renier}, \& {Cotter}}]{2021APh...12902579S}
{Spencer}, S., {Armstrong}, T., {Watson}, J., {et~al.} 2021, Astroparticle
  Physics, 129, 102579

\bibitem[{{Vinogradov}(2012)}]{2012NIMPA.695..247V}
{Vinogradov}, S. 2012, Nuclear Instruments and Methods in Physics Research A,
  695, 247

\bibitem[{Virtanen {et~al.}(2020)Virtanen, Gommers, Oliphant, Haberland, Reddy,
  Cournapeau, Burovski, Peterson, Weckesser, Bright, {van der Walt}, Brett,
  Wilson, Millman, Mayorov, Nelson, Jones, Kern, Larson, Carey, Polat, Feng,
  Moore, {VanderPlas}, Laxalde, Perktold, Cimrman, Henriksen, Quintero, Harris,
  Archibald, Ribeiro, Pedregosa, {van Mulbregt}, \& {SciPy 1.0
  Contributors}}]{2020SciPy-NMeth}
Virtanen, P., Gommers, R., Oliphant, T.~E., {et~al.} 2020, Nature Methods, 17,
  261

\end{thebibliography}
%

\end{document}